\documentclass[a4paper,11pt]{article}
\pdfoutput=1 
\usepackage{jheppub} 
\usepackage[utf8]{inputenc}
\usepackage{slashed}
\usepackage{amsfonts}
\usepackage{amsmath}
\usepackage{amssymb}
\usepackage{bbm}
\usepackage[vcentermath]{youngtab}
\usepackage[dvipsnames]{xcolor}
\usepackage[T1]{fontenc} 

\newcommand{\beq}{\begin{eqnarray}}
\newcommand{\eeq}{\end{eqnarray}}

\newcommand{\bmp}{\noindent\begin{minipage}{16cm}}
\newcommand{\emp}{\end{minipage}\vskip 7mm}

\newcommand{\SU}{\mbox{SU}}

\newcommand{\SP}{\mbox{Sp}}
\newcommand{\UU}{\mbox{U}}

\title{Addressing Six Standard Model Problems with Technically Natural Higgs Models}

\author[a]{Martin Rosenlyst}

\affiliation[a]{Rudolf Peierls Centre for Theoretical Physics, University of Oxford, 1 Keble Road, Oxford OX1 3NP, United Kingdom\\}

\emailAdd{martin.jorgensen@physics.ox.ac.uk}

\abstract{We aim to study the potential of the recently proposed model framework, called Technically Natural Higgs (TNH), in addressing six fundamental problems in particle physics and cosmology. These questions encompass the electroweak (EW) naturalness problem, neutrino mass generation, nature of the inflaton, the matter-antimatter asymmetry problem, origin of dark matter (DM) and the strong CP problem. We investigate various solutions within the TNH framework for three inflation scenarios --- Higgs, Starobinsky and scale-independent inflation. In the minimal TNH model, the Higgs is a mixture of an elementary and a composite state, with a compositeness scale far exceeding the EW scale. Traditionally, this has required an unnatural small vacuum misalignment, but in the TNH framework a novel mechanism enables a technically natural large compositeness scale, even up to the Planck scale. In this model framework, we demonstrate that a scale-invariant version of the minimal TNH model, featuring a special energy scale of around $ \mathcal{O}(10^{12}) $~GeV, loop-induced by the inflaton, simultaneously yields a technically natural 125-GeV Higgs boson, scotogenic neutrinos, a scale-invariant inflaton and a QCD axion DM candidate. These components dynamically generate the Planck scale and collectively have the potential to address all six open questions. } 

\begin{document} 
\maketitle
\flushbottom

\section{Introduction}
\label{sec: Introduction}

The primary objective of this paper is to address as many problems in particle physics and cosmology as possible through the introduction of a new model framework --- referred to as Technically Natural Higgs (TNH) in this study. This framework has recently been proposed in Ref.~\cite{Rosenlyst:2021tdr}. For the minimal TNH model, the Higgs boson is a mixture of an elementary and a composite state, with a compositeness scale much larger than the electroweak (EW) scale. Usually, achieving a large compositeness scale for a composite Higgs requires an artificially small vacuum misalignment. However, in the TNH framework, a novel mechanism enables a technically natural large compositeness scale, extending up to the Planck scale. As a result, this mechanism dynamically triggers EW symmetry breaking and fermion mass generation, including the neutrinos, in a technically natural way.

The Standard Model (SM) is currently the most successful theory of particle physics, accurately describing almost all known physical phenomena involving three of the four fundamental forces (excluding gravity) and elementary particles. Despite its impressive experimental predictions, the SM fails to adequately explain several fundamental phenomena in nature, leaving a number of questions unanswered. In this paper, we aim to address six of these outstanding puzzles using the recently proposed TNH model framework, namely \begin{itemize}
	\item[1)] the EW naturalness problem,
	\item[2)] neutrino masses and their flavour mixing,
	\item[3)] the inflationary paradigm,
	\item[4)] the matter-antimatter asymmetry problem,
	\item[5)] dark matter (DM), 
	\item[6)] the strong CP problem. 
\end{itemize}  Furthermore, we will briefly explore the prospects and challenges of developing a quantum gravity theory and resolving the cosmological constant problem within this framework.

An inherent challenge in constructing models beyond the SM (BSM) that can explain these open questions is the requirement for unnatural fine-tuning of the model parameters. Although the individual problems can sometimes be remedied with natural or technically natural fine-tuning, addressing several issues simultaneously often requires significant unnatural fine-tuning. In Ref.~\cite{Rosenlyst:2021tdr}, it is demonstrated how the minimal TNH model alleviates the EW naturalness problem and generates neutrino masses in a technically natural manner. Given the difficulty in constructing fully natural models, our objective in this paper is to resolve as many of the aforementioned problems as possible by introducing the TNH model framework with minimal unnatural fine-tuning.

\section{Introduction of six open questions in fundamental physics}
\label{sec: Introduction of the six open questions}

Now, we will present the six aforementioned problems and briefly discuss how they can potentially be addressed within the context of the TNH model framework.  \\

\textbf{The EW naturalness problem:} Our first inquiry seeks a fundamental origin of the Higgs sector of the SM to alleviate the SM naturalness problem. A dynamical origin of the Higgs sector and EW symmetry breaking (EWSB) is a possible solution of this problem. In Composite Higgs (CH) models~\cite{Kaplan:1983fs,Dugan:1984hq}, the Higgs boson arises as a composite pseudo-Nambu-Goldstone boson (pNGB) from a spontaneously broken global symmetry upon a compositeness scale $ \sim 4\pi f $, where $ f $ is the Goldstone-boson decay constant of the composite sector.\footnote{In general, composite Higgs models were first proposed in Refs.~\cite{Terazawa:1976xx,Terazawa:1979pj}, but in these models the Higgs boson is not identified as a light pNGB candidate. } It can, therefore, be naturally light with respect to the compositeness scale and other composite resonances in the model --- in good agreement with current observations. In this work, we will further assume the global symmetry of the Higgs sector arises as the chiral symmetries $ G $ of fermions of an underlying confining four-dimensional gauge-fermion theory. The new strong interactions dynamically break these global symmetries $ G $ to $ H $, while the interactions with the SM fields break the symmetries further, triggering the correct EWSB pattern dynamically.

In the CH framework, generating fermion masses is challenging. Approaches like Extended Technicolor (ETC)~\cite{Dimopoulos:1979es}, Partial Compositeness (PC)~\cite{Kaplan:1991dc}, Fundamental Partial Compositeness (FPC)~\cite{Sannino:2016sfx,Cacciapaglia:2017cdi} and Partially Composite Higgs (PCH)~\cite{Galloway:2016fuo,Agugliaro:2016clv} have been proposed but may face problems, like new naturalness issues, dangerous flavour changing neutral currents (FCNCs) or instability of the Higgs vacuum. These concerns can be eased with a high compositeness scale. The TNH framework suggests achieving this via a novel mechanism involving a softly breaking of a global $ \mathbb{Z}_2 $ symmetry by technically natural small vacuum misalignment. We focus on the PCH approach for simplicity and present the minimal partially composite two-Higgs scheme in Section~\ref{sec: A Concrete Partially Composite Higgs Model}, where the Higgs is a mixture of a composite and an elementary state, transforming odd under a global $ \mathbb{Z}_2 $ symmetry. Therefore, this model allows a large compositeness scale due to a technically natural small misalignment. An analysis in Ref.~\cite{Alanne:2017ymh} finds that PCH models without this mechanism suffer from low vacuum instability scales in renormalization group (RG) running. However, applying a similar analysis to the TNH model in Section~\ref{sec: RG analysis and vacuum stability} reveals that the proposed mechanism supports a large vacuum instability scale, effectively addressing this issue. \\

\textbf{Neutrino masses and their flavour mixing:} Regarding the second challenge, the only neutral fermions, neutrinos, have much smaller masses than charged fermions. In the early realizations of the Yukawa couplings~\cite{Weinberg:1967tq}, they were even thought to be massless, until the discovery of oscillations~\cite{Pontecorvo:1967fh} convinced the scientific community that they must carry a mass, although very small. The simplest solution to this puzzle is the seesaw mechanism~\cite{Minkowski:1977sc,Mohapatra:1979ia,Yanagida:1980xy}, based on the existence of very heavy new states that couple to neutrinos and the Higgs boson via large couplings. This requires a typical new scale $\Lambda_{\rm seesaw} \approx 10^{12}$~GeV. In Section~\ref{sec: Loop-Induced Neutrino Masses}, we explore the possibility of realizing the one-loop radiative seesaw mechanism by introducing three right-handed neutrinos (RHNs) and a second elementary $ \SU(2)_{\rm L} $ doublet to the partially composite two-Higgs scheme. Except for the new $ \SU(2)_{\rm L} $ scalar doublet is $ \mathbb{Z}_2 $--even and not odd, this setup is analogous to the one in the traditional scotogenic model~\cite{Ma:2006km}, arguably the simplest model of radiative neutrino masses. Additionally, in Section~\ref{sec: RG analysis and vacuum stability}, we perform a RG analysis for all couplings in this TNH model, incorporating scotogenic neutrino masses for different compositeness scales. This analysis demonstrates that the Higgs vacuum could be stable or metastable, and the SM naturalness problem can be alleviated with a large compositeness scale ranging from about $10^9$~GeV to the Planck scale. \\

\textbf{The inflationary paradigm:} For the third concern, cosmic inflation tackles horizon, flatness problems and generates density perturbations as seeds for today's large-scale structures~\cite{Hawking:1982cz,Starobinsky:1982ee,Guth:1982ec,Bardeen:1983qw}. This framework also accurately predicts inflationary parameters~\cite{Mukhanov:1981xt}, strongly aligned with Planck collaboration's Cosmic Microwave Background (CMB) anisotropy measurements~\cite{Planck:2018vyg,Planck:2018jri}. In Section~\ref{sec: The inflationary paradigm}, we examine three inflation scenarios with a Higgs doublet: \begin{itemize}
	\item[i)] Higgs Inflation~\cite{Bezrukov:2007ep},
	\item[ii)] Starobinsky $ R^2 $ Inflation~\cite{Starobinsky:1980te},
	\item[iii)] Scale-Invariant Inflation~\cite{Ferreira:2016wem,Shaposhnikov:2008xi,Shaposhnikov:2008xb,Blas:2011ac,Garcia-Bellido:2011kqb,Ferreira:2016vsc}. 
\end{itemize} However, the unitarity issue is important in defining the validity of these inflationary mechanisms. In the case of the original Higgs inflation, the traditional setup requires a large non-minimal coupling to the Ricci scalar, $ \alpha_H \sim 10^4 $, that leads to a cutoff scale far below the Planck scale, $ \Lambda \lesssim m_P/\alpha_H \ll m_P $~\cite{Burgess:2009ea,Barbon:2009ya,Burgess:2010zq,Lerner:2009na,Park:2018kst}, resulting in the unitarity problem. Various attempts have been proposed to treat this issue, for example by  introducing the Higgs near-criticality~\cite{Hamada:2014iga,Bezrukov:2014bra,Hamada:2014wna}, assuming a field-dependent vacuum expectation value~\cite{Bezrukov:2010jz} or adding new degrees of freedom~\cite{Giudice:2010ka,Barbon:2015fla,Giudice:2014toa,Ema:2017rqn,Gorbunov:2018llf}. In the case of the Starobinsky $ R^2 $ inflation model, the General Relativity is extended to include $ R^2 $ curvature squared corrections to the Einstein-Hilbert action. In this scenario, this dynamics can unitarize the Higgs inflation up to the Planck scale~\cite{Kehagias:2013mya,Giudice:2014toa}. However, according to Ref.~\cite{Ferreira:2021ctx}, the Higgs/$ R^2 $ inflation model introduce significant corrections to the SM Higgs mass, leading to an unacceptable hierarchy problem. The observed Higgs mass requires an extremely unnatural cancellation with the bare mass term with accuracy much greater than $ 1 $ part in $ 10^{13} $ for the pure Higgs inflation and $ 10^8 $ for Starobinsky inflation, smaller but still unnatural large. Therefore, they considered the case where inflation is driven by
the $ R^2 $ term with a scale-invariant extension. In this scenario, the hierarchy problem is reduced further such that we need a cancellation with the accuracy $ 1 $ part in $ 10^{4} $, which is still large. 

In Section~\ref{sec: The inflationary paradigm}, we will see that within the specific TNH model we are examining, the elementary Higgs doublets can possess arbitrary large bare masses, while the mass of the SM-like Higgs boson can be naturally fixed at its observed value of $125$~GeV. This effectively resolves the hierarchy issues associated with the three inflation scenarios. Another reason for considering scale-invariant inflation is its potential to dynamically explain the origin of the Planck mass through a new scalar field $\phi$ (identified as the inflaton) that breaks scale invariance spontaneously, obtaining a VEV. Additionally, we can assume that the masses of heavy fermions like the RHNs result from introducing Yukawa couplings between these fermions and $\phi $. \\

\textbf{The matter-antimatter asymmetry problem:} The fourth task aims to explain the baryon asymmetry of the Universe (BAU), quantified as the cosmic baryon-to-photon ratio $ \eta_B^{\rm obs}\simeq 6.1\times 10^{-10}$~\cite{ParticleDataGroup:2016lqr,Planck:2015fie}. The SM does not provide a solution to this question. An attractive method for generating the BAU dynamically is leptogenesis~\cite{Fukugita:1986hr}. In the context of standard thermal leptogenesis, closely tied to the type-I seesaw mechanism, small neutrino masses are achieved by introducing a minimum of two heavy RHNs with large Majorana masses. These heavy neutrinos emerge from interactions in the thermal environment, with their CP-violating out-of-equilibrium decays creating a primordial lepton asymmetry. EW sphaleron processes subsequently convert this lepton asymmetry into a baryon asymmetry.

Thermal leptogenesis in the scotogenic model has been explored extensively in previous studies. E.~Ma introduced it initially in Ref.~\cite{Ma:2006fn}, with further examination in subsequent works like Refs.~\cite{Kashiwase:2012xd,Kashiwase:2013uy,Racker:2013lua,Clarke:2015hta,Hugle:2018qbw}. More comprehensive summaries of leptogenesis studies in the scotogenic model are available, such as Ref.~\cite{Cai:2017jrq}. In Section~\ref{sec: Leptogenesis}, we will demonstrate that the minimal TNH model with scotogenic neutrino masses could potentially account for the observed BAU within the aforementioned inflation scenarios, all without requiring extra ingredients to the model.  \\

\textbf{Dark matter:} The fifth question pertains to the enigma of DM, constituting 85\% of the total mass today, yet unaccounted for by any SM particle. In Section~\ref{sec: Dark matter}, we briefly consider the possibility of composite super-heavy DM (SHDM) candidates. These SHDM particles acquire their relic density non-thermally by the weakness of the gravitational interactions~\cite{Chung:1998zb,Kuzmin:1998uv,Kuzmin:1998kk,Kolb:1998ki,Chung:1999ve,Kuzmin:1999zk,Chung:2001cb}. Indeed, this non-thermal gravitational production of the DM abundance during inflation is the only experimentally verified DM production mechanism, since the observed CMB fluctuations have exactly the same origin~\cite{Kannike:2016jfs}. Considering unverified alternatives, SHDM emerges as a natural DM candidate within this model. 

Extending the spectrum of new strongly interacting fermions with only one SM-singlet Dirac fermion, we can identify a DM candidate as a composite complex pNGB, $ \Phi $. This candidate, studied in Ref.~\cite{Dietrich:2006cm,Luty:2008vs,Ryttov:2008xe,Frandsen:2011kt,Alanne:2018xli,Rosenlyst:2021elz}, is protected by a global $\UU(1)$ symmetry and attains a mass of the order of the inflaton mass ($\sim \mathcal{O}(f)$). This SHDM mass lies within $1\times 10^{13}\text{GeV}\ll m_\Phi \lesssim 1\times 10^{17}\text{GeV}$ for Higgs inflation, $m_\Phi \sim 9\times 10^{9}~\text{GeV}$ for $ R^2 $ inflation and approximately $\mathcal{O}(10^{11})$~GeV in scale-invariant inflation scenario. \\

\textbf{The strong CP problem:} Answering the final question about the strong CP problem, we will explore a potential QCD axion candidate within this model. Many BSM models introduce Axion-Like-Particles (ALPs), which are pNGBs, arising from the spontaneous breaking of a global $ \UU(1) $ symmetry. In this way, these ALPs are naturally light particles, with a mass $ m_a $ much smaller than the symmetry breaking scale $ f_a $ (the ALP decay constant), and couple weakly to
the SM particles which scales as $ 1/f_a $. This ALP mass is typically generated by some new strong dynamics breaking the $ \UU(1) $ symmetry explicitly at some lower scale $ \Lambda_a $ compared to the scale $ f_a $ (i.e. $ \Lambda_a \ll f_a $), resulting in $ m_a \propto \Lambda_a^2 /f $. In the case with QCD, such an $ \UU(1) $ symmetry can be exact at the classical level but broken at the quantum level by a colour anomaly, leading to a spontaneous breaking of this symmetry at some high energy scale. This symmetry
is known as Peccei-Quinn (PQ) symmetry, $ \UU(1)_{\rm PQ} $, after its proposers~\cite{Peccei:1977hh,Peccei:1977ur} and such a broken symmetry results in an ALP with the mass $ m_a \propto \Lambda_{\rm QCD}^2/f_a $, called the QCD axion. Thus, this setup is a possible solution to the strong CP problem~\cite{Peccei:1977hh,Weinberg:1977ma,Wilczek:1977pj}, addressing the last of the above open questions. 

In Section~\ref{sec: QCD axion dark matter}, we identify the angular component of the inflaton $ \phi $ in the scale-invariant inflation scenario as a candidate for an ALP. This complex scalar inflaton $ \phi $ is charged under a global lepton $\UU(1)$ symmetry, spontaneously broken by its VEV ($v_\phi \sim 10^{15}$~GeV), resulting in its angular part $ a $ will be identified as a pNGB. To resolve the strong CP problem, we introduce a new chiral pair, $ \SU(3)_{\rm C} $ colored triplets $ Q_{L,R} $, with lepton numbers. Their coupling to $ \phi $ via Yukawa interactions leads to that this lepton $ \UU(1) $ symmetry becomes colour-anomalous, establishing it as a Peccei-Quinn (PQ) symmetry. Consequently, $a$ evolves into a QCD axion with $ f_a = v_\phi \sim 10^{15} $~GeV. In accordance with Ref.~\cite{Eroncel:2022vjg}, a standard misalignment mechanism could overproduce DM if the initial misalignment angle is around unity. However, for a misalignment angle between $0.01$ and $0.1$, this axion effectively constitutes cold DM in the Universe, simultaneously resolving the strong CP problem. Thus, via scale-invariant inflation, the enigmas of DM and the strong CP problem potentially find resolution through this QCD axion DM candidate with the mass $ m_a \sim 6\times 10^{-9} $~eV, albeit at the expense of some fine-tuning of the initial misalignment angle. \\

Throughout this paper, we will demonstrate that a concrete scale-invariant TNH model with a particular energy scale of around $ \mathcal{O}(10^{12}) $~GeV, loop-induced by the inflaton, may be able to answer the six aforementioned open questions in fundamental physics, while also dynamically generating the Planck scale. Finally, in Section~\ref{sec: Other open questions in fundamental physics}, we briefly discuss the possibilities and challenges associated with constructing a quantum gravity theory and resolving the cosmological constant problem within the context of this concrete scale-invariant model. Finally, we provide a summary and draw our conclusions in Section~\ref{sec: Summary}.

\section{The minimal technically natural Higgs model}
\label{sec: A Concrete Partially Composite Higgs Model}

In the following, we focus on the minimal TNH model, proposed in Ref.~\cite{Rosenlyst:2021tdr}. In this model, the concrete $\SU(6)/\SP(6)$ model~\cite{Cai:2018tet}, explored in Refs.~\cite{Cai:2019cow,Rosenlyst:2020znn,Cai:2020bhd,Cheng:2020dum,Cacciapaglia:2020psm}, is extended with one elementary $\mathbb{Z}_2$--odd Higgs doublet, $ H $. We assume that four strongly interacting Weyl fermions (hyper-fermions) are arranged in two $\SU(2)_{\rm L}$ doublets, $\Psi_1 \equiv (\psi_{1},\psi_{2})^T$ and $\Psi_2 \equiv (\psi_{5},\psi_{6})^T$, and two in $\SU(2)_{\rm L}$ singlets, $\psi_{3,4}$, with hypercharges $ \mp 1/2 $. In the minimal scenario, we assume that these six Weyl hyper-fermions are in fundamental representation of a new strongly interacting hypercolor (HC) gauge group $ \rm G_{\rm HC} = \rm SU(2)_{\rm HC} $ or $ \rm Sp(2N)_{\rm HC} $, which is the pseudo-real representation. For simplicity, we choose the minimal gauge group $ \SU(2)_{\rm HC} $ from now on. We can then construct an $ \SU(6) $ flavour multiplet by arranging the six Weyl hyper-fermions into a $ \SU(6) $ vector $ \Psi \equiv (\psi^1,\psi^2,\psi^3,\psi^4,\psi^5,\psi^6)^T $. We have listed in Table~\ref{tab:fermionssu6sp6} the representations of the gauge groups and parity under the $ \mathbb{Z}_2 $ symmetry of the fermions and the elementary weak doublet in the model. Therefore, differently from the SM, we assume that $ H $ is odd under the $ \mathbb{Z}_2 $ symmetry of the composite sector, leading to that the Higgs boson arise as a mixture between the $ \mathbb{Z}_2 $--odd composite pNGB from the spontaneously global symmetry breaking $ \SU(6)\rightarrow \SP(6) $ and the elementary weak doublet, $ H $. 

\begin{table}[tb]
\begin{center}
{\renewcommand{\arraystretch}{1.4}
\begin{tabular}{c|c|c|c|c|c}
\hline\hline
    & $\mbox{G}_{\rm HC} $ & $\mbox{SU(3)}_{\rm C} $  & $\mbox{SU(2)}_{\rm L} $ & $\mbox{U(1)}_{\rm Y} $ & $ \mathbb{Z}_2 $ \\
    \hline\hline
 $ \Psi_1\equiv (\psi_1,\psi_2)^T $  & $\Box$ & $ \textbf{1}$&  $\Box$ &  $0$ &$ +1 $  \\ 
  $ \psi_3 $  & $\Box$ & $\textbf{1}$&  $\textbf{1}$&  $-1/2$ &$ +1 $  \\ 
  $ \psi_4 $  & $\Box$ & $\textbf{1}$&  $\textbf{1}$&  $+1/2$ &$ +1 $  \\
 $ \Psi_2\equiv (\psi_5,\psi_6)^T $  & $\Box$ & $ \textbf{1}$&  $\Box$ &  $0$ &$ -1 $  \\
\hline\hline
 $ H $  & $\textbf{1}$ & $ \textbf{1}$&  $\Box$ &  $+1/2$ &$ -1 $  \\
\hline\hline
\end{tabular} }
\end{center}
\caption{The hyper-fermions in the SU(6)/Sp(6) template model and the elementary Higgs doublet, $ H $, labelled with their representations of $ \rm G_{\rm HC} \otimes SU(3)_C \otimes SU(2)_L \otimes U(1)_Y $ and parity under the $ {\mathbb{Z}_2} $ symmetry. } \label{tab:fermionssu6sp6}
\end{table}

Furthermore, via new Yukawa interactions between the strongly interacting hyper-fermions and $ H $, the VEV generated by the vacuum misalignment in the composite sector can be transferred to the neutral CP-even component of $ H $, leading to a VEV $ v $ of it. Based on that, the SM-fermions can achieve their masses via ordinary Yukawa couplings to $ H $. On the other hand, the EW gauge bosons obtain masses from both the VEV of the elementary Higgs doublet and the vacuum misalignment in the composite sector such that the EW scale is set by \beq \label{EW scale}
v_{\rm EW}^2 = v^2 + f^2 \sin^2\theta \,, 
\eeq where $ \theta $ ($ \pi/2 \leq \theta \leq \pi $) parameterises the vacuum misalignment. At $ \sin\theta =0 $ ($ \theta=\pi $) EW and $ \mathbb{Z}_2 $ symmetries are unbroken, while at $ \sin\theta =1 $ ($ \theta=\pi/2 $) the condensate is purely $ \SU(2)_{\rm L} $ and the $ \mathbb{Z}_2 $ symmetry is broken (this limit is commonly referred Bosonic Technicolor, proposed and explored in Refs.~\cite{Simmons:1988fu,Kagan:1990az,Kagan:1991gh,Carone:1992rh,Carone:1993xc}). This concrete PCH model example is thus technically natural even though that the vacuum misalignment angle $ \theta $ is very close to $ \pi $ since the $ \mathbb{Z}_2 $ symmetry is restored for $ \theta=\pi $. In this section, we consider this complete model in detail. 

\subsection{The condensate and pNGBs}

When the hyper-fermions confine, this results in the chiral symmetry breaking $ \SU(6)\rightarrow \text{Sp}(6) $ and the hyper-fermions develop a non-trivial and antisymmetric vacuum condnesate~\cite{Galloway:2010bp}
\beq \label{eq: condensate Phi sector}
\langle \Psi^I_{\alpha,a}\Psi^J_{\beta,b}\rangle\epsilon^{\alpha\beta}\epsilon^{ab}\sim f^3 E^{IJ}_{\rm CH}\,, \eeq where $ \alpha,\beta $ are spinor indices, $ a,b $ are HC indices, and $ I,J $ are flavour indices. We will suppress the contractions of these indices for simplicity. The vacuum of the composite sector, giving rise to the VEV of the neutral CP-even component $ \Phi_{\rm odd}^0 $ of the $ \mathbb{Z}_2 $--odd composite doublet, by misalignment, can be written as~\cite{Galloway:2010bp} \beq \label{EW vacuum matrix}
E_{\text{CH}}=\begin{pmatrix} i\sigma_2 & 0 &0\\ 0 &-i\sigma_2 c_\theta& -\mathbbm{1}_2 s_\theta\\0& \mathbbm{1}_2 s_\theta & i\sigma_2 c_\theta \end{pmatrix}\,,
\eeq where $ \sigma_2 $ is the second Pauli matrix and from now on we use the definitions $ s_x \equiv \sin x $, $ c_x\equiv \cos x $ and $ t_x \equiv \tan x $. The chiral symmetry breaking $ \SU(6)\rightarrow \text{Sp}(6) $ results in 14 pNGBs, $ \pi_a $ with $ a=1,...,14 $, and thus 14 $ \SU(6) $ broken generators, $ X_a $, correspond to the vacuum $ E_{\rm CH} $. The Goldstone bosons around the CH vacuum, $ E_{\text{CH}} $ in Eq.~(\ref{EW vacuum matrix}), are parametrized as \beq \label{eq: GB matrix}
\Sigma(x)=\exp\left[\frac{2\sqrt{2}i}{f}\pi_a(x) X_a\right]E_{\text{CH}}
\eeq with the decay constant $ f $ of them. 

\begin{table}[tb]
\begin{center}
{\renewcommand{\arraystretch}{1.4}
\begin{tabular}{c|c|c}
\hline\hline
  $\displaystyle \mathrm{G}_0/\mathrm{H}_0$  & $\begin{array}{c} \mathbb{Z}_2\mbox{--odd} \\ \mbox{pNGBs} \end{array}$ & $\begin{array}{c} \mathbb{Z}_2\mbox{--even} \\ \mbox{pNGBs} \end{array}$ \\
\hline\hline
  $\begin{array}{c} \Phi_{\rm even} = (2,1/2)_+ \\ \eta' = (1,0)_+ \end{array}$  &  $\begin{array}{c} \Phi_{\rm odd} = (2,1/2)_- \\ \Delta = (3,0)_- \\  \varphi^0 = (1,0)_- \end{array}$  & $\eta = (1,0)_+ $ 
 \\ 
\hline\hline
\end{tabular} }\end{center}
\caption{The pNGBs in the SU(6)/Sp(6) template model in the EW unbroken vacuum ($ s_\theta=0 $) labelled with their $ (\rm SU(2)_L, U(1)_Y)_{\mathbb{Z}_2}$ quantum numbers.} \label{tab:su6sp6}
\end{table}

In this model, we recognize a $ \mathbb{Z}_2 $ symmetry generated by a $ \SU(6) $ matrix, represented as $ P = \text{Diag}(1\,,1\,,1\,,1\,,-1\,,-1) $. This symmetry remains intact when the EW gauge symmetry is unbroken, i.e., when $ s_\theta=0 $. Within this context, the fields classified as $ \mathbb{Z}_2\mbox{--odd} $ are the composite pNGBs: $ \Phi_{\rm odd}^0, \ (\Phi_{\rm odd}^0)^*, \ \Phi_{\rm odd}^{\pm}, \ \Delta^0, \  \Delta^\pm $ and $  \varphi^0 $. We have listed in Table~\ref{tab:su6sp6} the quantum numbers for the EW unbroken vacuum ($ s_\theta =0  $) and $ \mathbb{Z}_2 $ parity of the pNGBs divided into the groupings: the $ \mathbb{Z}_2 $--even pNGBs in the minimal $ G_0/H_0=\SU(4)/\SP(4) $ CH subset~\cite{Galloway:2010bp}, and the additional $ \mathbb{Z}_2 $--odd and --even pNGBs in the rest of the $ \SU(6)/\SP(6) $ subset. In addition, we denote the neutral components of the composite weak doublets, $\Phi_{\rm even,odd}$, as follows
\beq \label{Eq: composite doublet neutral components}
\Phi_{\rm even}^{0}\equiv\frac{\phi_R-i\phi_I}{\sqrt{2}}\,,\quad\quad \Phi_{\rm odd}^{0}\equiv\frac{h-i\pi^3}{\sqrt{2}}\,,
\eeq  while the elementary $ \mathbb{Z}_2 $--odd doublet is written as \begin{equation} \begin{aligned}\label{Eq: elementary doublet}
H=\frac{1}{\sqrt{2}}\begin{pmatrix}\sigma_h-i\pi_h^3\\ -(\pi_h^2 +i\pi_h^1) \end{pmatrix}\,.
\end{aligned}\end{equation}

\subsection{The chiral Lagrangian and the effective potential} 
\label{section III}
 
The underlying Lagrangian of this specific PCH model describing the new strong sector and the elementary doublet can be written as~\cite{Alanne:2017ymh} \begin{equation} \begin{aligned}\label{Eq: fundamentalLagrangian}
\mathcal{L}_{\rm PCH}=&\Psi^\dagger i\gamma^\mu D_\mu \Psi + D_\mu H^\dagger D^\mu H - m_H^2 H^\dagger H-\lambda_H (H^\dagger H)^2 \\ & - \left(\frac{1}{2}\Psi^T M \Psi + y_U H_\alpha (\Psi^T P^\alpha \Psi)+ y_D \widetilde{H}_\alpha (\Psi^T \widetilde{P}^\alpha \Psi)+\rm h.c.\right)
\end{aligned}\end{equation} with $ \widetilde{H}\equiv\epsilon H^* $. For the Yukawa interactions $ y_{U,D} $, we have constructed the spurions \begin{equation} \begin{aligned}
2P^1_{ij}&=\delta_{i5}\delta_{j3}-\delta_{i3}\delta_{j5}\,, \quad 2P^2_{ij}=\delta_{i6}\delta_{j3}-\delta_{i3}\delta_{j6}\,, \\
2\widetilde{P}^1_{ij}&=\delta_{i5}\delta_{j4}-\delta_{i4}\delta_{j5}\,, \quad 2\widetilde{P}^2_{ij}=\delta_{i6}\delta_{j4}-\delta_{i4}\delta_{j6}\,.
\end{aligned}\end{equation} Finally, we have introudced the vector-like masses for the new hyper-fermions (preserving the EW gauge and $ \mathbb{Z}_2 $ symmetries) via the matrix \begin{equation}\begin{aligned} \label{Eq: vector-like masses}
M &= \begin{pmatrix} m_1i\sigma_2 & 0 &0 \\  0  & m_2i\sigma_2 & 0\\ 0 & 0 & -m_3i\sigma_2 \end{pmatrix}\,.
\end{aligned} \end{equation} Note that for $ m_1=m_2=m_3 $, the mass matrix is proportional to the EW-preserving vacuum in Eq.~(\ref{EW vacuum matrix}) with $ \theta=\pi $ ($ c_\theta =-1 $), akin to the minimal PCH model~\cite{Galloway:2016fuo,Agugliaro:2016clv}. This is not by chance, as it is indeed the hyper-fermion masses that determine the signs in the vacuum structure~\cite{Cacciapaglia:2020kgq}.

To introduce the SM-like Yukawa couplings, we need to demand that the left- and right-handed SM-quarks and -leptons transform as $ q_{L,i} \equiv (u_{L,i},d_{L,i})^T \rightarrow q_{L,i}$, $ l_{L,i} \equiv (\nu_{L,i},e_{L,i})^T \rightarrow l_{L,i}$, $ u_{R,i} \rightarrow -u_{R,i} $, $ d_{R,i}\rightarrow -d_{R,i} $ and $ e_{R,i}\rightarrow -e_{R,i} $ under the $ \mathbb{Z}_2 $ symmetry. Thus, the elementary $ \mathbb{Z}_2 $--odd scalar doublet, $ H $, can couple to the SM-fermions like the Higgs doublet in the SM with the Yukawa interactions preserving the $ \mathbb{Z}_2  $ symmetry:\begin{equation} \begin{aligned}\label{Eq: Higgs-Yukawa couplings}
\mathcal{L}_{\rm Y}=&-y_t^{ij} \overline{q}_{L,i} H u_{R,j} - y_b^{ij} \overline{q}_{L,i} \widetilde{H} d_{R,j}- y_e^{ij} \overline{l}_{L,i} \widetilde{H} e_{R,j} +\rm h.c. \,.
\end{aligned}\end{equation}
 
Below the condensation scale, $ \Lambda_{\rm HC}\sim 4\pi f $, Eq.~(\ref{Eq: fundamentalLagrangian}) yields the following leading-order effective potential \begin{equation} \begin{aligned}\label{Eq: leading order effective potential}
V_{\rm eff}^0 = & m_H^2 H^\dagger H+\lambda_H (H^\dagger H)^2 \\ & - 4 \pi f^3 Z \bigg(\frac{1}{2}\text{Tr}[M\Sigma]-y_U H_\alpha \text{Tr}[P^\alpha\Sigma]-y_D \widetilde{H}_\alpha \text{Tr}[\widetilde{P}^\alpha\Sigma]+\text{h.c.}\bigg)\,,
\end{aligned}\end{equation} where $ Z $ is a non-perturbative $ \mathcal{O}(1) $ constant that can be suggested by lattice simulations (e.g. $ Z\approx 1.5 $ in Ref.~\cite{Arthur:2016dir} for the $ \SU(2) $ gauge theory with two Dirac (four Weyl) hyper-fermions). To  next-to-leading order, the EW gauge interactions contribute to the effective potential, which can be written as \begin{equation}
\begin{aligned}  \label{Eq: gauge loop contributions} V_{\rm gauge}^{\rm 1-loop}&= C_g f^4 \bigg(\sum_{i=1}^3 g_L^2 \text{Tr}[T_L^i \Sigma T_L^{iT}\Sigma^\dagger]+g_Y^{2}\text{Tr}[T_R^3 \Sigma T_R^{3T} \Sigma^\dagger]\bigg) \\ &=- C_g f^4 \left(\frac{3g_L^2+g^{2}_Y}{2}c_\theta^2+\dots\right)\,, \end{aligned} \end{equation} 
where $ g_{L,Y} $ are the $ \SU(2)_{\rm L} $ and $ \UU(1)_{\rm Y} $ gauge couplings, $ C_g $ is an $ \mathcal{O}(1) $ form factor that can be computed on the lattice and the gauged generators embedded in the $ \SU(2)_{\rm L}\otimes \SU(2)_{\rm R} $ subgroup of the global symmetry group $\SU(6)$ are identified by the left- and right-handed generators \begin{equation}\begin{aligned}
T_{L}^i &= \begin{pmatrix} \sigma_i & 0 &0 \\  0  & 0 & 0\\ 0 & 0 & \sigma_i \end{pmatrix}\,, \quad \quad T_{R}^i =\begin{pmatrix}0  & 0 &0 \\  0  & -\sigma_i^T  & 0\\ 0 & 0 & 0 \end{pmatrix}
\end{aligned} \end{equation} with $ i=1,2,3 $ and $ \sigma_i $ denoting the Pauli matrices. This effective potential is at the one-loop level, and accordingly, the contribution is subleading comparable to the vector-like mass terms. However, we will include this contribution in the following numerical calculations, because these terms are essential for generating the smallness of the neutrino masses in this model, as discussed in Section~\ref{sec: Loop-Induced Neutrino Masses}.  

In the following, we briefly ignore the next-to-leading order potential contribution in Eq.~(\ref{Eq: gauge loop contributions}) from the EW gauge interactions for simplicity. By assuming $ \langle \sigma_h \rangle =v$ in Eq.~(\ref{Eq: elementary doublet}), the minimization of the effective Lagrangian in Eq.~(\ref{Eq: leading order effective potential}) yields the parameter expressions \begin{equation} \begin{aligned}\label{Eq: minimizing effective potential 2}
 m_{23}=-\frac{c_\theta m_\lambda^2 t_\beta^2}{8\pi Z f}\,,\quad \quad y_{UD}=\frac{t_\beta m_\lambda^2 }{4\sqrt{2}\pi Z f^2}
\end{aligned}\end{equation} with $ t_{\beta}\equiv v/(fs_\theta) $ and $ m_\lambda^2 \equiv m_H^2+\lambda_H v^2 $. Unlike the Higgs mechanism in the SM, the squared mass parameter $ m_H^2 $ of the the elementary Higgs doublet in Eq.~(\ref{Eq: leading order effective potential}) does not need to change the sign to trigger the EWSB. This is a further motivation for investigating this minimal TNH model.

According to the Higgs potential in Eq.~(\ref{Eq: leading order effective potential}), the neutral CP-even scalar mass matrix in the basis $ (\sigma_h,h) $ can be written as \begin{equation} \begin{aligned}\label{Eq: Higgs mass matrix}
M_{\rm Higgs}^2=m_\lambda^2\begin{pmatrix}1+\delta & -c_\theta t_\beta \\ -c_\theta t_\beta & t_\beta^2 \end{pmatrix}\,, 
\end{aligned}\end{equation} where $ \delta\equiv 2\lambda_H v^2/m_\lambda^2 $. The mass eigenstates are given in terms of the interaction eigenstates by\begin{equation} \begin{aligned}\label{Eq: Higgs states}
h_1=c_\alpha \sigma_h -s_\alpha h\,, \quad\quad h_2=s_\alpha \sigma_h+c_\alpha h 
\end{aligned}\end{equation} with \begin{equation} \begin{aligned}\label{Eq: alpha}
t_{2\alpha}=\frac{2t_\beta c_\theta}{1-t_\beta^2 +\delta}\,.
\end{aligned}\end{equation} The corresponding masses of these eigenstates are \begin{equation} \begin{aligned}\label{Eq: Higgs masses}
m_{h_{1,2}}^2 = \frac{m_\lambda^2}{2}\Big[1+t_\beta^2 +\delta &\pm (2c_\theta t_\beta s_{2\alpha}+(1-t_\beta^2+\delta)c_{2\alpha})\Big]\,.
\end{aligned}\end{equation}

\section{Loop-induced neutrino masses}
\label{sec: Loop-Induced Neutrino Masses}

\begin{figure}[t!]
	\centering
	\includegraphics[width=0.45\textwidth]{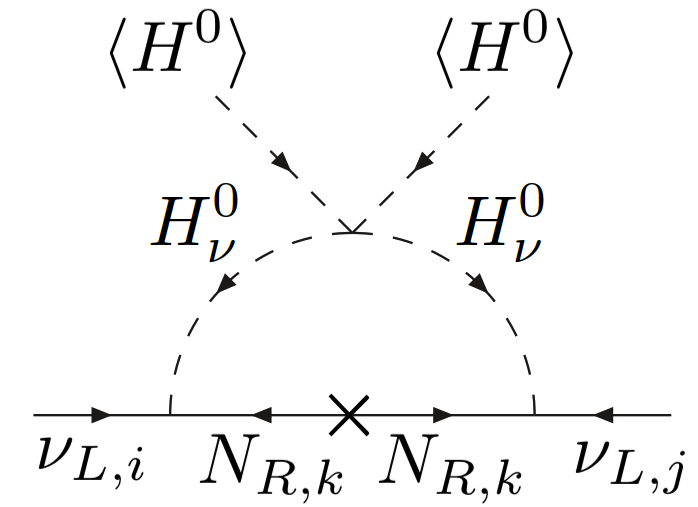} 
	\caption{One-loop radiative Majorana neutrino masses in this model, similar to the scotogenic model proposed in Ref.~\cite{Ma:2006km}.}
	\label{LoopDiagram}
\end{figure}

On the other hand, the generation of the small neutrino masses and their flavour mixing is an important subject. The simplest solution to this puzzle is the seesaw mechanism~\cite{Minkowski:1977sc,Mohapatra:1979ia,Yanagida:1980xy}, which requires a typical new scale $ \Lambda_{\rm seesaw}\approx 10^{12} $ GeV. In this section, we explore the possibility of realizing a one-loop radiative seesaw mechanism in this technically natural multi-Higgs scheme by considering the studies in Ref.~\cite{Cacciapaglia:2020psm}. However, the neutral component of a $ \mathbb{Z}_2 $--even doublet, not of a $ \mathbb{Z}_2 $--odd as in Ref.~\cite{Ma:2006km}, runs in the loop as in Figure~\ref{LoopDiagram}.

\subsection{The Lagrangian and scalar potential terms in the neutrino sector}  
 
To incorporate this mechanism, we need to add a new $ \mathbb{Z}_2 $--even elementary weak doublet with hypercharge $ +1/2 $,\begin{equation} \begin{aligned}\label{Eq: elementary doublet neutrino}
H_\nu=\frac{1}{\sqrt{2}}\begin{pmatrix}\sigma_R-i\sigma_I\\ -(\pi_\nu^2 +i\pi_\nu^1) \end{pmatrix}\,,
\end{aligned}\end{equation} and three right-handed neutrino SM-singlets, $ N_{R,i} $ ($ i=1,2,3 $) with Majorana masses, $ M_i $, transforming even under the global $ \mathbb{Z}_2 $ symmetry. 

With this assignment of parities under the $ \mathbb{Z}_2 $ symmetry of the fermions, Yukawa couplings between the elementary Higgs doublet $ H $ achieving the VEV $ v $ and the neutrinos are not allowed without explicitly breaking the $ \mathbb{Z}_2 $ symmetry. On the other hand, new fundamental Yukawa couplings with the new $ \mathbb{Z}_2 $--even doublet, $ H_\nu $, and the neutrino fields can be written as\begin{equation} \begin{aligned}\label{Eq: Higgs-Yukawa couplings with neutrinos}
\mathcal{L}_{\rm Y}\supset -h^{ij} \overline{l}_{L,i} H_\nu N_{R,j} +\rm h.c. \,.
\end{aligned}\end{equation} Furthermore, the new elementary $ \mathbb{Z}_2 $--even doublet, $ H_\nu $, is allowed to be coupled to the composite Higgs sector
by adding new fundamental Yukawa couplings between $ H_\nu $ and the $ \mathbb{Z}_2 $--even hyper-fermions to Eq.~(\ref{Eq: fundamentalLagrangian}): \begin{equation}\begin{aligned} \label{Eq: PCH Lagrangian with neutrinos Yukawa}
\mathcal{L}_{\rm PCH}\supset & - y_1 H_{\nu,\alpha} (\Psi^T P^\alpha_\nu \Psi) -y_2 \widetilde{H}_{\nu,\alpha} (\Psi^T \widetilde{P}^\alpha_\nu \Psi)+\rm h.c.\,,
\end{aligned} \end{equation} where we introduce the spurions \begin{equation} \begin{aligned}
2P^1_{\nu,ij}=\delta_{i1}\delta_{j3}-\delta_{i3}\delta_{j1}\,, \quad 2P^2_{\nu,ij}=\delta_{i2}\delta_{j3}-\delta_{i3}\delta_{j2}\,, \\
2\widetilde{P}^1_{\nu,ij}=\delta_{i1}\delta_{j4}-\delta_{i4}\delta_{j1}\,, \quad 2\widetilde{P}^2_{\nu,ij}=\delta_{i2}\delta_{j4}-\delta_{i4}\delta_{j2}\,. 
\end{aligned}\end{equation} 

Below the condensation scale, $ \Lambda_{\rm HC}\sim 4\pi f $, Eq.~(\ref{Eq: PCH Lagrangian with neutrinos Yukawa}) yields the following contributions to the effective potential in Eq.~(\ref{Eq: leading order effective potential}): \begin{equation} \begin{aligned}\label{Eq: leading order effective potential neutrino}
V_{\rm eff}^0 \supset \ & 4 \pi f^3 Z_\nu \left(y_1 H_{\nu,\alpha} \text{Tr}[P^\alpha_\nu\Sigma]+y_2 \widetilde{H}_{\nu,\alpha} \text{Tr}[\widetilde{P}^\alpha_\nu\Sigma]+\text{h.c.}\right)\,,
\end{aligned}\end{equation} where $ Z_\nu $ is a non-perturbative $ \mathcal{O}(1) $ constant. From now on, we assume that $ Z_\nu\equiv Z\approx 1.5 $~\footnote{According to the lattice calculations in Ref.~\cite{Arthur:2016dir}, $ Z\approx 1.5 $ for a $ \SU(2) $ gauge theory with two Dirac (four Weyl) hyper-fermions, which is only one Dirac fermion less compared to the concrete model considered in this paper.} and $ y_1=y_2 $ for simplicity.

Finally, we add all the allowed terms including the new doublet to the underlying Lagrangian in Eq.~(\ref{Eq: fundamentalLagrangian}) that conserves all the symmetries, which are given by \begin{equation}\begin{aligned} \label{Eq: PCH Lagrangian with neutrinos}
\mathcal{L}_{\rm PCH}\supset & D_\mu H_\nu^\dagger D_\mu H_\nu - m_{H_\nu}^2 H_\nu^\dagger H_\nu - \lambda_{H_\nu}(H_\nu^\dagger H_\nu)^2  - \lambda_{1}H^\dagger H H_\nu^\dagger H_\nu \\ & -\lambda_{2}H^\dagger H_{\nu} H_\nu^\dagger H-\bigg(\lambda_{3} (H^\dagger H_{\nu})^2 +\rm h.c.\bigg)\,. 
\end{aligned} \end{equation} In the following, we assume $ m_{H_\nu}\equiv m_H =f $ for avoiding naturalness problems. Moreover, we set $ \lambda_{1,2,3}=1 $ for simplicity, which has an insignificant influence on the RG running of $ \lambda_H $ since the components of $ H_{\nu} $ have large $ \mathcal{O}(f) $ masses and are ``integrated out'' in the following stability analysis. Note that the above potential contributions from the neutrino sector do not change the masses of the mass eigenstates $ h_{1,2} $ in Eq.~(\ref{Eq: Higgs masses}), where $ h_1 $ is still identified with the 125-GeV SM-Higgs boson. Furthermore, the new elementary doublet has no influence on the vacuum structure discussed in Section~\ref{sec: A Concrete Partially Composite Higgs Model}.

\subsection{The masses and mixing of the neutrinos}

In particular, the mixing mass matrices $ M_R^2 $
and $ M_I^2 $ in the bases $ (\sigma_R,\phi_R) $ and $ (\sigma_I,\phi_I) $, respectively, are generated by the potential of the neutral components of the elementary and composite $ \mathbb{Z}_2 $--even Higgs doublets in Eq.~(\ref{Eq: composite doublet neutral components}) and~(\ref{Eq: elementary doublet neutrino}), respectively, which are given by \begin{equation} \begin{aligned}\label{Eq: MR MI mass matrices} 
M_R^2 =& \widetilde{M}^2+\begin{pmatrix}2\lambda_3 v^2 & 0  \\ 0  & \frac{1}{2}C_g (3g_L^2+g_Y^2)f^2c_\theta \end{pmatrix}\,, \\
 M_I^2 =& \widetilde{M}^2+\begin{pmatrix}-2\lambda_3 v^2 & 0  \\ 0  & \frac{1}{2}C_g (8g_L^2+(g_Y^2-5 g_L^2)c_\theta)f^2 \end{pmatrix}
\end{aligned}\end{equation} with \begin{equation} \begin{aligned}
\widetilde{M}^2\equiv\begin{pmatrix}m_{H_\nu}^2 +\frac{1}{2}(\lambda_1+\lambda_2)v^2 & -4\sqrt{2}\pi Z y_{12} f^2 c_{\theta/2}  \\ -4\sqrt{2}\pi Z  y_{12} f^2 c_{\theta/2}  & 8\pi Zf m_1 +\frac{1}{2}m_\lambda^2 t_\beta^2 \end{pmatrix}\,, \nonumber
\end{aligned}\end{equation} where $ y_{12}\equiv y_1+y_2 $, $ t_{\beta}\equiv v/(fs_\theta) $ and $ m_\lambda^2 \equiv m_H^2+\lambda_H v^2 $. Therefore, a mass splitting is generated between the masses $ m_{R,I} $ of the mass eigenstates $ \widetilde{\sigma}_{R,I} $, consisting mostly of the neutral components $ \sigma_{R,I} $ in Eq.~(\ref{Eq: elementary doublet neutrino}), respectively. 

Assuming that the RHNs, $ N_{R,i} $, are not much heavier than the neutral components of the new $ \mathbb{Z}_2 $--even doublet, small Majorana masses of the left-handed neutrinos are generated by the loop diagram shown in Figure~\ref{LoopDiagram}, analog to the one in the traditional scotogenic model~\cite{Ma:2006km}. The loop diagram results in the mass expression~\cite{Ma:2006km}{\setlength{\thickmuskip}{.01\thickmuskip}
  \setlength{\medmuskip}{0.7\medmuskip}\begin{equation} \begin{aligned}\label{Eq: neutrino mass matrix formula}
m_\nu^{ij}=&\sum^3_{k=1}\frac{h^{ik}h^{jk}}{(4\pi)^2}M_k \bigg[\frac{m_{R}^2}{m_R^2-M_k^2}\ln\left(\frac{m_R^2}{M_k^2}\right)-\frac{m_{I}^2}{m_I^2-M_k^2}\ln\left(\frac{m_I^2}{M_k^2}\right)\bigg]\equiv \sum^3_{k=1}h^{ik}h^{jk}\Xi_{\nu,k} \,,
\end{aligned}\end{equation}}where $ M_i $ denotes the masses of the RHNs, $ N_{R,i} $. To obtain non-zero neutrino masses, we need a mass splitting between the masses $ m_{R,I} $ of the mass eigenstates $ \widetilde{\sigma}_{R,I} $, respectively, given by Eq.~(\ref{Eq: MR MI mass matrices}). 

By ignoring the EW gauge loop potential contributions in Eq.~(\ref{Eq: MR MI mass matrices}), i.e. $ g_{L,Y}=0 $, the mass splitting between the mass eigenstates $ \widetilde{\sigma}_{R,I} $ will only depend on the $ \lambda_3 v^2 $ term, which is negligible relative to their masses since $ m_{R,I}\sim f \gg v $. In this scenario, either the neutrino Yukawa couplings $ h^{ij} $ in Eq.~(\ref{Eq: Higgs-Yukawa couplings with neutrinos}) or the quartic coupling $ \lambda_3 $ in Eq.~(\ref{Eq: PCH Lagrangian with neutrinos}) needs to be non-perturbative large ($ h^{ij},\lambda_3 > 4\pi $) to achieve large enough neutrino masses. When the gauge interactions are turned on, a mass splitting will be generated in the order of $ f $ between the composite states $ \phi_{R,I} $, resulting in a more significant mass splitting of $ \widetilde{\sigma}_{R,I} $ via the Yukawa couplings, $ y_{1,2} $, in Eq.~(\ref{Eq: PCH Lagrangian with neutrinos Yukawa}). As shown in the following, this mass splitting may be enough to generate large enough neutrino masses with perturbative couplings. In this model, the neutrino masses are thus dynamically loop-induced by the composite dynamics via the EW gauge and Yukawa interactions. 

Before we present the numerical calculations, we need to define the neutrino mass matrix in Eq.~(\ref{Eq: neutrino mass matrix formula}), which can be diagonalized as \begin{equation}
\begin{aligned} \label{eq: neutrino masses and mixing}
	m_\nu^{\rm Diag} &= U_{\rm PMNS}^T m_\nu U_{\rm PMNS}={\rm Diag}(m_{\nu_1},m_{\nu_2},m_{\nu_3}),  
\end{aligned}
 \end{equation} where $ m_{\nu_{i}} $ with $ i=1,2,3 $ are the left-handed neutrino masses. The matrix $ U_{\rm PMNS}=U U_m $ is the Pontecorvo--Maki--Nakagawa--Sakata (PMNS) matrix, where $ U_m =\rm Diag(1,e^{i\phi_1/2},e^{i\phi_2/2}) $ encoding the Majorana phases and the matrix $ U $ is parametrized as \beq
\begin{pmatrix}c_{12}c_{13}&s_{12}c_{13}&s_{13}e^{-i\delta}\\-s_{12}c_{23}-c_{12}s_{23}s_{13}e^{i\delta} & c_{12}c_{23}-s_{12}s_{23}s_{13}e^{i\delta}  & s_{23}c_{13} \\ s_{12}s_{23}-c_{12}c_{23}s_{13}e^{i\delta}& -c_{12}s_{23}-s_{12}c_{23}s_{13}e^{i\delta}&c_{23}c_{13}\end{pmatrix} \nonumber
\eeq with the mixing angles $ s_{ij}\equiv \sin\theta_{ij},\,c_{ij}\equiv \cos\theta_{ij} $ and the Dirac phase $ \delta $. In this paper, we assume that the Majorana and Dirac phases are vanishing ($ \phi_{1,2}=0 $ and $ \delta=0 $), but it is possible to add them without significant changes of our conclusions. 

In following section, we fit to the best-fit experimental values for the mass-squared differences ($ \Delta m_{ij}\equiv m_{\nu_i}^2-m_{\nu_j}^2 $) and mixing angles ($ s_{ij}\equiv \sin\theta_{ij} $), which are given in Ref.~\cite{Esteban:2018azc} for both normal hierarchy (NH) of the neutrinos, i.e. $ m_{\nu_1}<m_{\nu_2}<m_{\nu_3} $, and inverted hierarchy (IH), i.e. $ m_{\nu_3}<m_{\nu_1}<m_{\nu_2} $. Finally, we also include the upper bound on the sum of the neutrino masses coming from cosmology. The most reliable bound is from the Planck collaboration~\cite{Ade:2015xua}, \beq
	\sum_i m_{\nu_i} \lesssim 0.23 \rm \ eV\,. \label{eq: neutrino mass sum bound}
\eeq

In the following calculations, we have chosen NH of the neutrinos, where $ m_{\nu_1}=0.0010 $ eV, leading to $ m_{\nu_2}=0.0087 $ eV, $ m_{\nu_3}=0.0500 $ eV, and therefore, the bound in Eq.~(\ref{eq: neutrino mass sum bound}) is fulfilled. If we choose IH or another value of $ m_{\nu_1} $, the following conclusions will not change significantly. Furthermore, we assume that the RHN masses are $  M_i/2 = m_H = m_{H_\nu} = f $ GeV and the vector-like mass of the hyper-fermion doublet $ \Psi_1 $ is $ m_1 = m_{2,3}/2 $, because those values give rise to the most favorable results.

\section{Vacuum stability analysis across diverse $\Lambda_{\rm HC}$}
\label{sec: RG analysis and vacuum stability}

As concluded in Ref.~\cite{Alanne:2017ymh}, PCH models may suffer from a low vacuum instability scale, which is the energy scale below the Planck scale where the quartic coupling, e.g. $ \lambda_H $ in Eq.~(\ref{Eq: fundamentalLagrangian}), runs to negative values. Such an instability originates from the fact that the top-Yukawa coupling of the elementary interaction eigenstate, $ \sigma_h $, is enhanced compared to the SM-Higgs~\cite{Carone:2012cd} by \begin{equation} \begin{aligned}\label{Eq: enhancement of the top Yukawa coupling}
y_t=y_t^{\rm SM}/s_\beta\,,
\end{aligned}\end{equation} where $ y_t $ and $ y_t^{\rm SM} $ are the enhanced and SM top-Yukawa couplings, respectively.

In Ref.~\cite{Rosenlyst:2021tdr}, we have performed a numerical analysis of the RG running of the quartic coupling $ \lambda_H $ in Eq.~(\ref{Eq: fundamentalLagrangian}) with the compositeness scale $ \Lambda_{\rm HC}=m_{P} $. Moreover, in this RG analysis, we assume that the top-Yukawa and the EW gauge couplings run as in the SM below the compositeness scale, since all the values of the couplings are SM-like after the condensation and we neglect the effects on the quartic coupling running during the confinement. We, therefore, use the SM RG equations up to three loops, listed in Ref.~\cite{Buttazzo:2013uya}. We assume further that no composite states other than the Higgs state affect the running of $ \lambda_H $ due to the fact that all these states are integrated out below $ f\approx 1.9\times 10^{17} $ GeV, because their masses are between $ f $ and $ \sim f t_\beta  $ for small $ t_\beta^{-2} $.

In the following, we will construct a scale-invariant TNH model with potential answers to questions raised in Section~\ref{sec: Introduction of the six open questions}, including the EW naturalness problem, the neutrino masses, inflation, matter-antimatter asymmetry, dark matter and the strong CP problem. However, this model necessitates a compositeness scale below the Planck scale. Hence, in the next section, we will perform an extensive RG analysis, encompassing different compositeness scales.

\subsection{The RG equations with scotogenic neutrinos}

In the subsequent discussion, we will employ the abbreviation: \begin{equation} \begin{aligned}\label{Eq: }
\mathcal{D}\equiv 16\pi^2\mu \frac{d}{d\mu}
\end{aligned}\end{equation} with $ \mu $ being the renormalization scale. Furthermore, we define $ N_F $ as the number of new $ \SU(2)_{\rm L} $ hyper-fermion doublets, $ \Psi_{i} $, that transform under the representation $ \mathcal{R} $ under the new strong gauge group, $ G_{\rm HC} $, and couple either to the $ \mathbb{Z}_2 $--odd and --even elementary Higgs doublets ($ H $ and $ H_\nu $) via Yukawa interactions ($ y_{U,D} $ and $ y_{1,2} $). 

Without the composite sector, the RG equations (RGEs) of the couplings have been calculated for the SM and its extended version including scotogenic neutrinos in Ref.~\cite{Bouchand:2012dx}. In the following, we write the RGEs of the couplings in our considered model that are valid above the compositeness scale, $ \Lambda_{\rm HC} $. Above $ \Lambda_{\rm HC} $, the RGEs of the scalar couplings in Eqs.~(\ref{Eq: fundamentalLagrangian}) and~(\ref{Eq: PCH Lagrangian with neutrinos}) can be written as \begin{equation} \begin{aligned}\label{Eq: beta functions 1}
\mathcal{D}\lambda_H=& 12\lambda_H^2 +2\lambda_1^2+2\lambda_1\lambda_2 +\lambda_2^2 +2\lambda_3^2+\frac{3}{8}(g_Y^4+3g_L^4+2g_Y^2 g_L^2)-6y_t^4\\&-3\lambda_H\left(g_Y^2+3g_L^2-4y_t^2\right)+4 d(\mathcal{R})(y_U^2+y_D^2)\lambda_H-d(\mathcal{R}) (y_U^4+y_D^4)\,,  \\
\mathcal{D}\lambda_{H_\nu}=& 12\lambda_{H_\nu}^2+2\lambda_1^2+2\lambda_1\lambda_2
+\lambda_2^2+2\lambda_3^2+\frac{3}{8}(g_Y^4+3g_L^4+2g_Y^2 g_L^2)-2\text{Tr}(h^\dagger hh^\dagger h)\\&-3\lambda_{H_\nu}\left(g_Y^2+3g_L^2-\frac{4}{3}\text{Tr}(h^\dagger h)\right)  +4 d(\mathcal{R})(y_1^2+y_2^2)\lambda_{H_\nu}-d(\mathcal{R}) (y_1^4+y_2^4)\,,\\
\mathcal{D}\lambda_1=& 4(\lambda_H+\lambda_{H_\nu})(3\lambda_1+\lambda_2)+4\lambda_1^2+2\lambda_2^2
+4\lambda_3^2+\frac{3}{4}(g_Y^4+3g_L^4-2g_Y^2 g_L^2)\\&-3\lambda_1\left(g_Y^2+3g_L^2-4y_t^2-\frac{4}{3}\text{Tr}(h^\dagger h)\right) \,, \\ 
\mathcal{D}\lambda_2=& 2 (\lambda_H+\lambda_{H_\nu})\lambda_2+8\lambda_1\lambda_2+4\lambda_2^2
+16\lambda_3^2-3\lambda_2\left(g_Y^2+3g_L^2-2y_t^2-\frac{2}{3}\text{Tr}(h^\dagger h)\right)\\ & +3g_Y^2g_L^2 \,,  \\
\mathcal{D}\lambda_3=& 4(\lambda_H+\lambda_{H_\nu})\lambda_3+8\lambda_1\lambda_3
+12\lambda_2\lambda_3-3\lambda_3\left(g_Y^2+3g_L^2-2y_t^2-\frac{2}{3}\text{Tr}(h^\dagger h)\right) \,,
\end{aligned}\end{equation} where $ d(\mathcal{R}) $ is the dimension of the representation $ \mathcal{R} $ and $ h $ is the Yukawa coupling matrix of the neutrinos defined in Eq.~(\ref{Eq: Higgs-Yukawa couplings with neutrinos}). Moreover, the RGEs of the gauge couplings in this concrete model are given by \begin{equation} \begin{aligned}\label{Eq: beta functions 2}
 \mathcal{D}g_{\rm HC}=&-\left(\frac{11}{3}C_2(\mathcal{A})-\frac{8}{3}N_F T(\mathcal{R})\right)g_{\rm HC}^3\,,\\
  \mathcal{D}g_{Y}=&\left(7+\frac{4}{3}N_F d(\mathcal{R})\left(4Y(Q_L)^2+\frac{1}{2}\right)\right)g_{Y}^3\,,\\
  \mathcal{D}g_{L}=&-\left(3-\frac{2}{3}N_F d(\mathcal{R})T(\mathcal{R})\right)g_{L}^3\,,\\
  \mathcal{D}g_{C}=&-7g_{C}^3\,,
\end{aligned}\end{equation} where $ T(\mathcal{R}) $ is the index of the representation $ \mathcal{R} $, $C_2(\mathcal{A}) $ is the quadratic Casimir of the adjoint representation, $ g_C $ is the strong gauge coupling and $ Y(\Psi_{i}) $ is the hypercharge of the $ \SU(2)_{\rm L} $ hyper-fermion doublets, $ \Psi_{i} $, which is zero in our case. Finally, the RGEs for the Yukawa couplings in Eqs.~(\ref{Eq: fundamentalLagrangian}),~(\ref{Eq: Higgs-Yukawa couplings}),~(\ref{Eq: Higgs-Yukawa couplings with neutrinos}) and~(\ref{Eq: PCH Lagrangian with neutrinos Yukawa}) can be written as \begingroup\allowdisplaybreaks \begin{align}\label{Eq: beta functions 3}
  \mathcal{D}y_{t}=&\left(\frac{9}{2}y_t^2-\frac{17}{12}g_Y^2-\frac{9}{4}g_L^2-8g_C^2+d(\mathcal{R})(y_U^2+y_D^2)\right)y_t\,,\nonumber\\
 \mathcal{D}y_{U}=&\left(-6C_2(\mathcal{R})g_{\rm HC}^2-\frac{9}{4}g_L^2-\frac{17}{12}g_Y^2+3y_t^2+\left(d(\mathcal{R})+\frac{3}{2}\right)y_U^2+d(\mathcal{R})y_D^2\right)y_U\,, \nonumber \\\mathcal{D}y_{D}=&\left(-6C_2(\mathcal{R})g_{\rm HC}^2-\frac{9}{4}g_L^2-\frac{5}{12}g_Y^2+3y_t^2+\left(d(\mathcal{R})+\frac{3}{2}\right)y_D^2+d(\mathcal{R})y_U^2\right)y_D\,,\\
 \mathcal{D}y_{1}=&\left(-6C_2(\mathcal{R})g_{\rm HC}^2-\frac{9}{4}g_L^2-\frac{17}{12}g_Y^2+3y_t^2+\left(d(\mathcal{R})+\frac{3}{2}\right)y_1^2+d(\mathcal{R})y_2^2\right)y_1\,,\nonumber\\\mathcal{D}y_{2}=&\left(-6C_2(\mathcal{R})g_{\rm HC}^2-\frac{9}{4}g_L^2-\frac{5}{12}g_Y^2+3y_t^2+\left(d(\mathcal{R})+\frac{3}{2}\right)y_2^2+d(\mathcal{R})y_1^2\right)y_2\,,\nonumber\\\mathcal{D}h=&h\left(\frac{3}{2}h^\dagger h+\text{Tr}(h^\dagger h) -\frac{3}{4}g_Y^2-\frac{9}{4}g_L^2\right)\,.\nonumber
\end{align}\endgroup 

As in the RG analysis in Ref.~\cite{Rosenlyst:2021tdr}, we neglect the effects on the quartic coupling running during the confinement and assume that the top-Yukawa and the EW gauge couplings run as in the SM below the compositeness scale, $ \Lambda_{\rm HC} $. In the following, we, therefore, use the SM RGEs up to three loops in Ref.~\cite{Buttazzo:2013uya} below $ \Lambda_{\rm HC} $. We assume further that no composite states other than the Higgs state affect the running of $ \lambda_H $ below $ \Lambda_{\rm HC}\sim f $, since all these states are integrated out below $ f $ due to the fact that their masses are between $ f $ and $ \sim f t_\beta  $ for small $ t_\beta^{-2} $. Finally, above $ \Lambda_{\rm HC} $, the RG running of the different couplings in this model are described by the RGEs in Appendix A in Ref.~\cite{Alanne:2017ymh} without neutrinos and the above RGEs with neutrinos, given by Eqs.~(\ref{Eq: beta functions 1})--(\ref{Eq: beta functions 3}). However, if the energy scale, where the above RGEs switch to the SM-like RGEs in the RG analysis, is lowered from $ \Lambda_{\rm HC}\approx 4\pi f $ to a value between $ f $ and $ \Lambda_{\rm HC} $, there are no significant changes of the following conclusions.

\subsection{Numerical results} 

\begin{figure}[t!]
	\centering
	\includegraphics[width=0.6\textwidth]{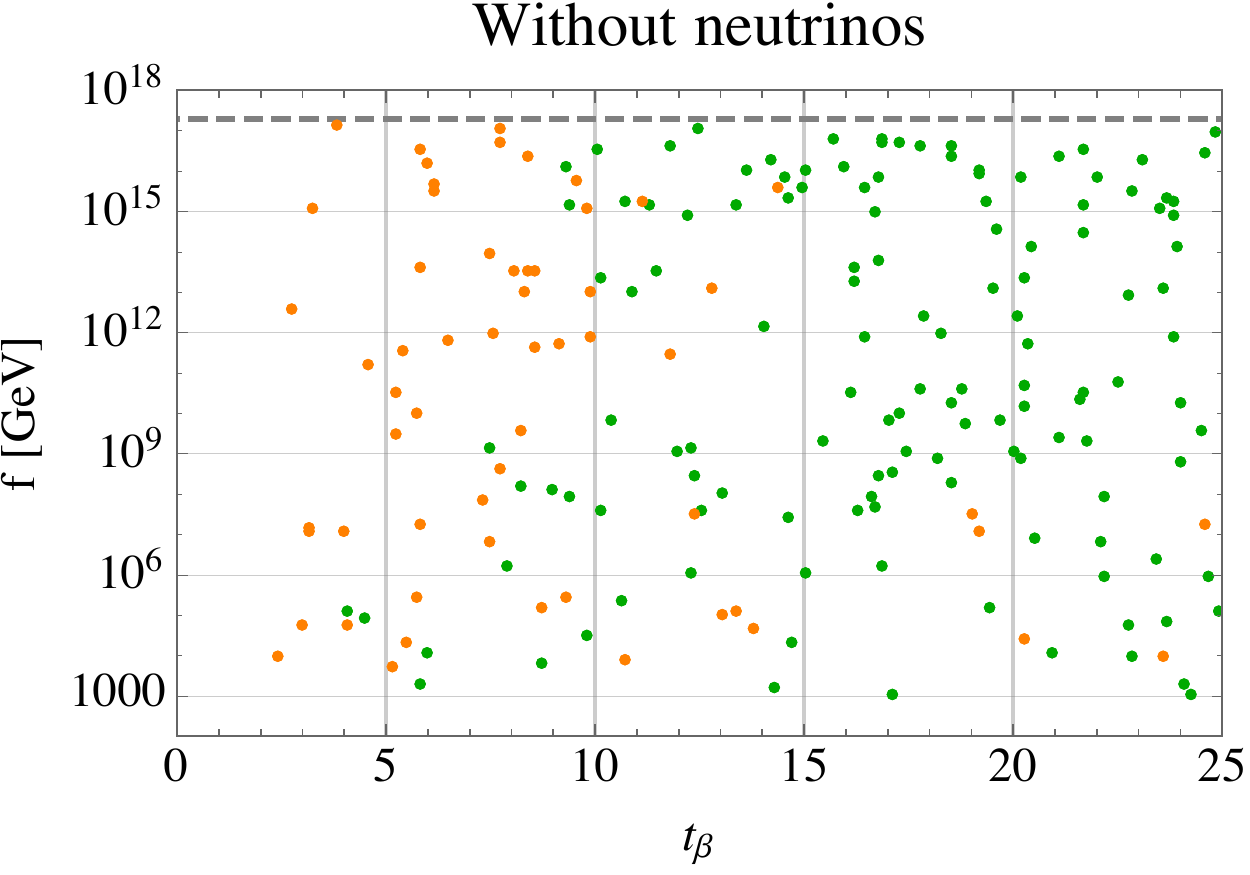} 
	\includegraphics[width=0.6\textwidth]{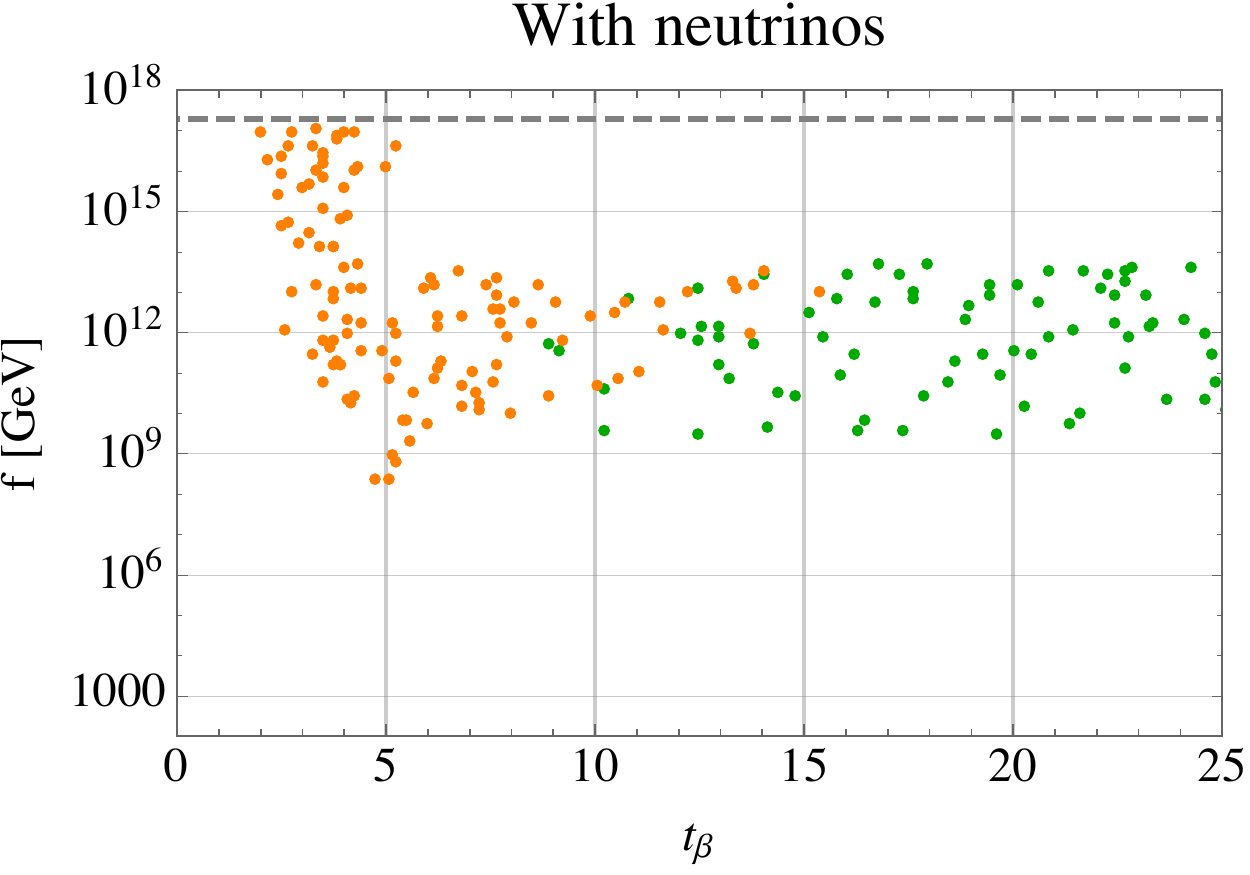} 
	\caption{The stability of the vacuum in the model examples without (upper panel) and with scotogenic neutrinos (lower panel) in the $ t_\beta $--$ f $ parameter space. The orange and green parameter points give rise to metastable and stable Higgs vacua, respectively. The top mass is in the $ 2\sigma $ band set by CMS, $ m_t=170.59\dots 173.65 $~GeV~\cite{CMS:2021jnp}. In this RG analysis, we have randomly scanned over the dimensionless variables in the ranges $0.1 \leq\lambda_{H_\nu},\lambda_{1,2,3}\leq 1 $, $ 0.5 \leq y_{1,2} \leq 5 $ and $ 0.01 \leq h^{ij} \leq 10 $, while $ m_H=m_{H_\nu}=M_i = f $ and $ m_{1,2,3}\approx 0.3 f $. }
	\label{fig: RG running without with neutrinos}
\end{figure}

By temporarily ignoring the neutrinos, the model features eight parameters relevant to our study: $ m_1 $, $ m_{23}\equiv m_2+m_3 $, $ y_{UD}\equiv y_U+y_D $, $ m_H $, $ \lambda_H $, $ s_\theta $, $ f $ and $ t_\beta $; and four constraint equations including the EW scale in Eq.~(\ref{EW scale}), the two vacuum conditions in Eq.~(\ref{Eq: minimizing effective potential 2}) and the Higgs mass in Eq.~(\ref{Eq: Higgs masses}). In this section, we set $ m_1=m_{23}/2 $ for simplicity (e.g. $ m_1=m_2=m_3 $). Furthermore, the SM naturalness problem is alleviated by a large $ m_H $. However, the vacuum quickly becomes unstable when we increase $ m_H $ above $ f $~\cite{Alanne:2017ymh}. Therefore, we assume $ m_H = f $ with a large $ f $, which requires a technically natural small $ s_\theta $. Thus, we can assume that $ f $ and $ t_\beta $ are free parameters. If the neutrinos are added, we also have following parameters: $ m_{H_\nu} $, $ M_i $, $ \lambda_{H_\nu} $, $ \lambda_{1,2,3} $, $ y_{1,2} $ and $ h^{ij} $. Firstly, we assume $ m_{H_\nu}=m_H = f $ for avoiding a new hierarchy problem. Secondly, the RHN masses are set to be $ M_i = f $, which makes it easier to introduce a scale invariance in the model when all the dimensionful fundamental parameters in the model are of the order of $ f $. Finally, the vector-like masses are given by  Eq.~(\ref{Eq: minimizing effective potential 2}) to be $ m_{1,2,3}\approx 0.3 f  $. Altogether, we have $ m_H=m_{H_\nu}=M_i = f $ and $ m_{1,2,3}\approx 0.3 f $. In the following RG analysis, we randomly scan over the dimensionless variables in the ranges $0.1 \leq\lambda_{H_\nu},\lambda_{1,2,3}\leq 1 $, $ 0.5 \leq y_{1,2} \leq 5 $ and $ 0.01 \leq h^{ij} \leq 10 $ in the $ t_\beta  $--$ f $ parameter space. Finally, we assume that the non-perturbative coefficient of the potential contribution from the EW gauge interactions in Eq.~(\ref{Eq: gauge loop contributions}) is in the interval $  0.5 \leq C_g \leq 5 $. 

\begin{figure}[t!]
	\centering
	\includegraphics[width=0.94\textwidth]{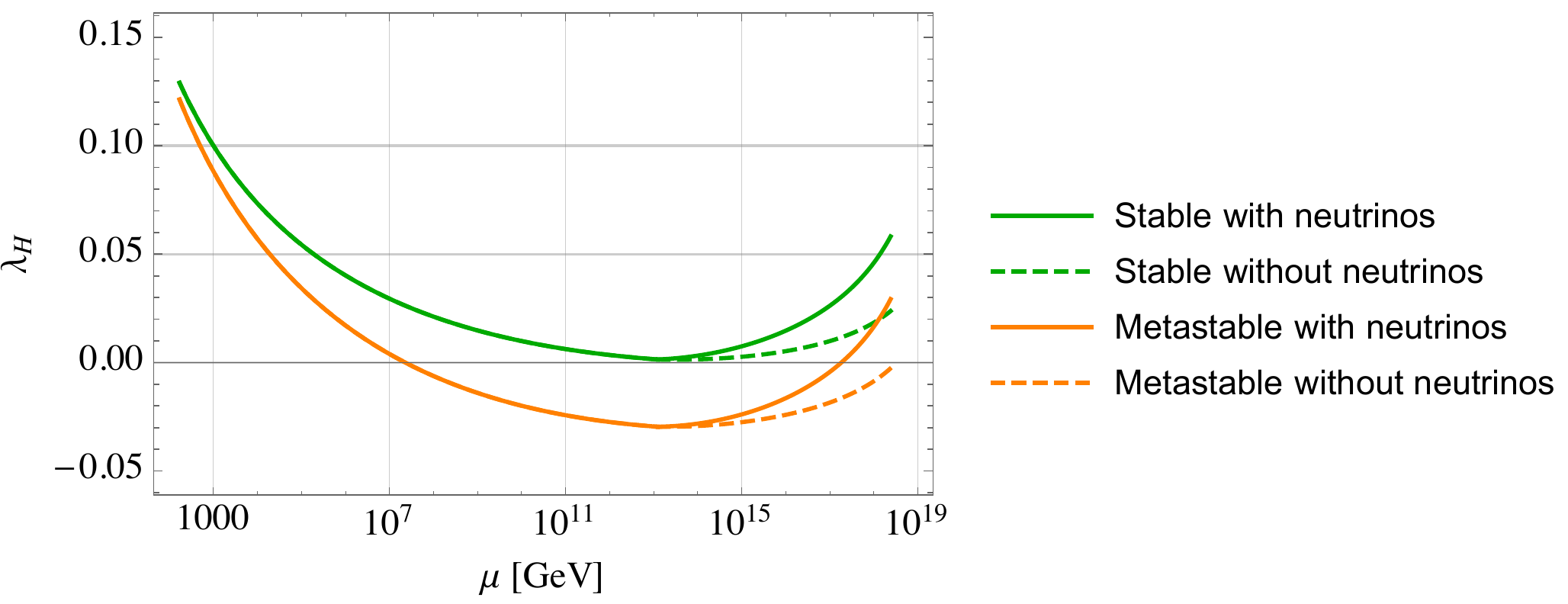} 
	\caption{Comparison of the RG running of the quartic coupling $ \lambda_H $ for a stable ($ t_\beta=20 $) and metastable vacuum ($ t_\beta=5 $) with and without induced neutrino masses. In these cases, $ m_H=f=10^{12} $~GeV and $ C_g=1 $, while $ m_{H_\nu}=f $, $ \lambda_{H_\nu}=0.1 $, $ y_{1,2}=1 $ and $ 0.28>\vert h_{ij} \vert >0.02 $ when loop-induced neutrino masses are included. The top mass is set to its central value $ 172.13 $~GeV set by CMS~\cite{CMS:2021jnp}. Finally, we assume the mass hierarchy $ M_{2,3}=10M_1=f $. }
	\label{fig: RG running examples}
\end{figure}

In Figure~\ref{fig: RG running without with neutrinos}, phase diagrams of the vacuum stability are depicted for the model examples without (upper panel) and with scotogenic neutrinos (lower panel) in terms of $ t_\beta $ and $ f $, where we assume that the top mass is within the $ 2\sigma $ band set by CMS: $ m_t=170.59\dots 173.65 $~GeV~\cite{CMS:2021jnp}. The orange and green parameter points represent, respectively, metastable and stable vacua in the $ t_\beta  $--$ f $ parameter space, while the gray dashed line represents the value of $ f\approx 1.9\times 10^{17} $~GeV giving rise to a compositeness scale at the Planck scale, $ \Lambda_{\rm HC} = m_{P} $. As shown in Figure~2 in Ref.~\cite{Rosenlyst:2021tdr}, for the special case with $ \Lambda_{\rm HC} = m_{P} $ and without neutrinos, we need a relative large $ t_\beta \gtrsim 13 $ to acquire a stable vacuum. On the other hand, according to upper panel in Figure~\ref{fig: RG running without with neutrinos} without neutrinos, it seems that the required minimum magnitude of $ t_\beta $ providing a stable vacuum decreases slightly for decreasing decay constant $ f $. However, a smaller decay constant $ f $ introduces a new naturalness problem, since $ m_H =f $. In the case with neutrinos, we also need a large $ t_\beta $, but we cannot achieve a stable vacuum with $ \Lambda_{\rm HC}=m_{P} $ as discussed in Section~\ref{sec: Loop-Induced Neutrino Masses}. However, according to the lower panel in Figure~\ref{fig: RG running without with neutrinos},  the vacuum can be stable if the decay constant is somewhere in the interval $ 10^{9}~\text{GeV} \lesssim f \lesssim 10^{14}$~GeV for $ t_\beta \gtrsim 9 $. In what follows, it will become clear that inflation along with a technically natural Higgs and loop-induced neutrino masses can be realized in various scenarios when the decay constant has a value in the range $ 10^{9}~\text{GeV} \lesssim f \lesssim 10^{14}$~GeV. 

Figure~\ref{fig: RG running examples} shows some examples of the RG running of the quartic coupling $ \lambda_H $ for $ f=10^{12} $~GeV without and with neutrinos, where the vacuum is either stable ($ t_\beta =20 $) or metastable ($ t_\beta=5 $). We have set $ C_g=1 $ for all four examples, while $ \lambda_{H_\nu}=0.1 $ and $ y_{1,2}=1 $ for the two examples including loop-induced neutrino masses. For each of the two neutrino cases, there exists one positive, real solution of the neutrino Yukawa coupling constants in Eq.~(\ref{Eq: Higgs-Yukawa couplings with neutrinos}) with maximal number of zeroes by using Eq.~(\ref{Eq: neutrino mass matrix formula}). For these two examples with stable and metastable vacuum, the matrices of the neutrino Yukawa couplings are, respectively, given by \begin{equation} \begin{aligned}\label{Eq: neutrino mass matrix examples}
&h^{ij}=\begin{pmatrix}0.14 & 0.09 & 0.02 \\ 0 & 0.18 & 0.22 \\ 0 & 0 & 0.26 \end{pmatrix}\Rightarrow \textcolor{Green}{\text{ Stable vacuum}} \\
&h^{ij}=\begin{pmatrix}0.15 & 0.10 & 0.02 \\ 0 & 0.20 & 0.24 \\ 0 & 0 & 0.28 \end{pmatrix}\Rightarrow \textcolor{orange}{\text{ Metastable vacuum}}
\end{aligned}\end{equation} For these examples, we have assumed that the RHN masses are $ M_{2,3}=10 M_1 = f $, which is an interesting mass hierarchy with respect to the leptogenesis in this concrete model, discussed later in Section~\ref{sec: Leptogenesis}. If $ \mathcal{O}(0.1)f\lesssim M_{i} \lesssim  \mathcal{O}(10)f $, a change of the mass hierarchy of them has no significant effect on the RG analysis and does not change the order of magnitude of the Yukawa couplings $ h^{ij} $. \\

In the following section, we will explore the diverse avenues for realizing cosmic inflation within the TNH framework, both in scenarios without and with the inclusion of neutrinos.

\section{The inflationary paradigm}
\label{sec: The inflationary paradigm}

In this section, we partially follow the naturalness analysis performed in Ref.~\cite{Ferreira:2021ctx} of three different inflation mechanisms: original Higgs inflation~\cite{Bezrukov:2007ep}, Starobinsky $ R^2 $ inflation~\cite{Starobinsky:1980te} and scale-invariant inflation~\cite{Ferreira:2016wem,Shaposhnikov:2008xi,Shaposhnikov:2008xb,Blas:2011ac,Garcia-Bellido:2011kqb,Ferreira:2016vsc}. However, according to Ref.~\cite{Ferreira:2021ctx}, the Higgs/$ R^2 $ inflation model introduce significant corrections to the SM Higgs mass, leading to an unacceptable hierarchy problem. The observed Higgs mass requires an extremely unnatural cancellation with the bare mass term with accuracy much greater than $ 1 $ part in $ 10^{13} $ for the original Higgs inflation and $ 10^8 $ for Starobinsky inflation, smaller but still unnatural large. Therefore, they considered the case where inflation is driven by
the $ R^2 $ term with a scale-invariant extension. In this case, hierarchy problem is reduced further such that we need a cancellation with the accuracy $ 1 $ part in $ 10^{4} $, which is still large. 

We will show that each of these inflation approaches can be realized in the TNH framework both without and with neutrinos, while the above mentioned hierarchy problem can be completely remedied. In what follows, we will first introduce the three inflation scenarios, then we will analyze the parameter space where these scenarios can be realized in the TNH framework, and finally we will calculate the reheating temperature after inflation of a concrete scale-invariant TNH model and give the results of the reheating after the Higgs/$ R^2 $ inflation from Ref.~\cite{Aoki:2022dzd}.

\subsection{Mixed Higgs/$ R^2 $ inflation}
\label{sec: Mixed R2/Higgs inflation}

Now, we will introduce the scale dependent Higgs/$ R^2 $ inflation model~\cite{He:2018gyf}. When the $ \mathbb{Z}_2 $--odd elementary Higgs doublet, $ H $ in Eq.~(\ref{Eq: elementary doublet}), is present, in the unitary gauge, the Starobinsky inflation model is given by the action\begin{equation}\begin{aligned} 
S=\int \sqrt{-g}\left(\frac{1}{2}g^{\mu\nu}\partial_\mu \sigma_H \partial_\nu \sigma_H - \frac{1}{6f_0^2} R^2 - \frac{\alpha_H}{12} \sigma_H^2 R +\frac{m_P^2}{2} R-\frac{\lambda_H}{4}\left(\sigma_H^2-v^2\right)^2  \right)\,, \label{eq: mixed R2/Higgs inflation action}
\end{aligned} \end{equation} where $ \sigma_h $ is the CP-even neutral component of $ H $ in Eq.~(\ref{Eq: elementary doublet}), $ R(g) $ is the Ricci scalar of the metric $ g_{\mu\nu} $ and $ m_P $ is the Planck mass. 

The $ R^2 $ term in the above action contains an additional scalar degree of freedom~\cite{Whitt:1984pd,Hindawi:1995an} due to the fact that it involves fourth order derivatives. These fourth order derivatives can be reduced to second order by introducing the auxiliary field $ \chi $, called scaleron. With this reduction, the action can now rewritten as \begin{equation}\begin{aligned} 
S=\int \sqrt{-g}\left(\frac{1}{2}g^{\mu\nu}\partial_\mu \sigma_H \partial_\nu \sigma_H + \frac{m_P^2}{2} R \Omega^2 -\frac{\lambda_\chi}{4}\chi^4-\frac{\lambda_H}{4}\left(\sigma_H^2-v^2\right)^2 \right)\,,
\end{aligned} \end{equation} where 
\begin{equation}\begin{aligned} 
\lambda_\chi &\equiv\frac{1}{24} f_0^2 \alpha_\chi^2 \\ 
\Omega^2 & \equiv 1-\frac{\alpha_H}{6m_P^2}\sigma_H^2 - \frac{\alpha_\chi}{6m_P^2}\chi^2\equiv \exp \left(\frac{2\omega}{\sqrt{6}m_P}\right) \,.
\end{aligned} \end{equation} If the $ R^2 $ term dominates the inflation, the field $ \omega $ will be identified as the inflaton. 

According to Refs.~\cite{Hill:2020oaj,Hill:2021psc}, the anomalous couplings to the Ricci scalar, proportional to $ \alpha_H $ and $ \alpha_\chi $, provide contact terms
in the effective Lagrangian below the Planck scale, which is driven by single graviton exchange. In Refs.~\cite{Hill:2020oaj,Hill:2021psc}, they show that these contact terms are obtained from a Weyl transformation where the action transforms from the Jordan frame to the Einstein frame, while for the effective Lagrangian the Jordan frame is already the Einstein frame. By including these contact terms or performing a Weyl transformation, the action is given by \begin{equation}\begin{aligned} 
S=\int \sqrt{-g}\bigg(&\frac{1}{2}\Omega^{-2} g^{\mu\nu}\partial_\mu \sigma_H \partial_\nu \sigma_H + \frac{1}{2}g^{\mu\nu}\partial_\mu \omega\partial_\nu \omega +\frac{m_P^2}{2} R -\frac{\lambda_\chi}{4}\Omega^{-4}\chi^4 \\ &-\frac{\lambda_H}{4}\Omega^{-4}\left(\sigma_H^2-v^2\right)^2\bigg)\,,
\end{aligned} \end{equation} where the quartic term of the auxiliary scaleron field $ \chi $ can be rewritten as \begin{equation}  \begin{aligned} \frac{\lambda_\chi}{4}\Omega^{-4}\chi^4   = \frac{3}{8}m_P^4 f_0^2 \left[1- \left(1-\frac{\alpha_H}{6m_P^2}\sigma_H^2\right)\exp\left(-\frac{2\omega}{\sqrt{6}m_P}\right)\right]^2 \,. \end{aligned} \end{equation}  

\subsection{Scale-independent $ R^2 $ inflation}
\label{sec: Scale independent R2 inflation}

The above mixed Higgs/$ R^2 $ model can be made scale-invariant by replacing the Planck mass term in the action in Eq.~(\ref{eq: mixed R2/Higgs inflation action}) as follows\begin{equation}\begin{aligned} 
\frac{m_P^2}{2}R\rightarrow \frac{1}{2}g^{\mu\nu}\partial_\mu\phi \partial_\nu \phi -\frac{\lambda_\phi}{4}\phi^4 -\frac{1}{12}\alpha_\phi \phi^2 R\,,
\end{aligned} \end{equation} where the Planck scale is induced via the new scalar field $ \phi $ achieving a VEV due to a spontaneous breaking of the scale invariance~\cite{Ferreira:2016wem,Shaposhnikov:2008xi,Shaposhnikov:2008xb,Blas:2011ac,Garcia-Bellido:2011kqb,Ferreira:2016vsc}. Then a scale-invariant action can be written as {\setlength{\thickmuskip}{.1\thickmuskip}
  \setlength{\medmuskip}{1.1\medmuskip}\begin{equation}\begin{aligned} 
S=\int \sqrt{-g}\bigg(&\frac{1}{2}g^{\mu\nu}\left(\partial_\mu \phi \partial_\nu \phi +\partial_\mu \sigma_H \partial_\nu \sigma_H\right) +\frac{m_P^2}{2} R\Omega^2-\frac{\lambda_\chi}{4}\chi^4-\frac{\lambda_\phi}{4}\phi^4-\frac{\lambda_H}{4}\sigma_H^4 \bigg)\,, \label{eq: SI action 1}
\end{aligned} \end{equation}} where \begin{equation}\begin{aligned} 
\Omega^2\equiv -\frac{\alpha_\chi}{6m_P^2}\chi^2-\frac{\alpha_\phi}{6m_P^2}\phi^2-\frac{\alpha_H}{6m_P^2}\sigma_H^2\,.
\end{aligned} \end{equation} This action is invariant under a global Weyl (scale) symmetry, where the fields transforms under this symmetry as follows \begin{equation}\begin{aligned} 
&g_{\mu\nu}(x)\rightarrow \Omega^{-2}g_{\mu\nu}(x)\,,\quad \quad g^{\mu\nu}(x)\rightarrow \Omega^{2}g^{\mu\nu}(x)\,, \quad \quad \sqrt{-g} \rightarrow \Omega^{-4}\sqrt{-g}\,, \quad\quad \\
& R(x) \rightarrow \Omega^4 R(x)\,, \quad\quad\quad\quad  \lbrace\sigma_H(x),\phi(x),\chi(x)\rbrace \rightarrow \Omega \lbrace\sigma_H(x),\phi(x),\chi(x)\rbrace\,.
\end{aligned} \end{equation}

Due to the fact that this model is invariant under this Weyl symmetry transformation, there exists a conserved Weyl current associated to this scale symmetry. This model undergoes inertial spontaneous scale symmetry breaking~\cite{Ferreira:2018itt}, where there appears a massless dilaton, $ \sigma (x) $. This symmetry transformation leads to a conserved Noether current $  K_\mu = \partial_\mu K $, where the kernal is given by \begin{equation}\begin{aligned} 
K=\frac{1}{6}\left[(1-\alpha_\phi)\phi^2+(1-\alpha_H)\sigma_H^2-\alpha_\chi \chi^2\right]\,.
\end{aligned} \end{equation} It is convenient to change the variables as follows \begin{equation}\begin{aligned} 
g_{\mu\nu}=e^{2\sigma(x)/f_\sigma}\hat{g}_{\mu\nu}\,, \quad \quad (\sigma_H,\phi,\chi) = e^{-\sigma(x)/f_\sigma}(\hat{\sigma}_H,\hat{\phi},\hat{\chi})\,,
\end{aligned} \end{equation} where $ f_\sigma $ has dimension of energy. Then, we can rewrite the kernal as \begin{equation}\begin{aligned} 
K=\frac{1}{6}\left[(1-\alpha_\phi)\hat{\phi}^2+(1-\alpha_H)\hat{\sigma}_H^2-\alpha_\chi \hat{\chi}^2\right]e^{-2\sigma /f_\sigma} \equiv \overline{K}e^{-2\sigma/f_\sigma}\,,
\end{aligned} \end{equation} where $ \overline{K} $ is a constant. We can view the inertial symmetry breaking~\cite{Ferreira:2018itt} as a red shifting of the dilaton to a constant VEV $ v_\sigma $, i.e. $ \sigma(x) \rightarrow v_\sigma + \hat{\sigma}(x)  $, leading to $ K\rightarrow \overline{K}e^{-2v_\sigma/f_\sigma}\equiv m_P^2 $. Thus, we can write the action as follows\begin{equation}\begin{aligned} 
S=\int \sqrt{-g}\bigg(& \frac{1}{2} g^{\mu\nu}\left(\partial_\mu \hat{\sigma} \partial_\nu \hat{\sigma}+\partial_\mu \hat{\phi} \partial_\nu \hat{\phi}+\partial_\mu \hat{\sigma}_H \partial_\nu \hat{\sigma}_H\right)  +\frac{m_P^2}{2}R\hat{\Omega}^2\\ &-
\frac{\lambda_\chi}{4}\hat{\chi}^4-\frac{\lambda_\phi}{4}\hat{\phi}^4-\frac{\lambda_H}{4}\hat{\sigma}_H^4  \bigg)\,, \label{eq: SI action 2}
\end{aligned} \end{equation} where \begin{equation}\begin{aligned} 
\hat{\Omega}^2= 1-\frac{1}{6m_P^2}\hat{\phi}^2-\frac{1}{6m_P^2}\hat{\sigma}_H^2\equiv \exp\left(\frac{2\theta}{\sqrt{6}m_P}\right)\,.
\end{aligned} \end{equation}

Finally, due to single graviton exchanges, the anomalous couplings in Eq.~(\ref{eq: SI action 1}) are replaced by contact terms in the effective Lagrangian (equivalent to a Weyl transformation~\cite{Hill:2020oaj}). The final form of the scale-invariant inflation action is then given by \begin{equation}\begin{aligned} 
S=\int \sqrt{-g}\bigg(&\frac{m_P^2}{2} R +\frac{1}{2} g^{\mu\nu}\partial_\mu \theta \partial_\nu \theta  +\frac{1}{2}\hat{\Omega}^{-2}g^{\mu\nu}\left(\partial_\mu \hat{\sigma} \partial_\nu \hat{\sigma}+\partial_\mu \hat{\phi} \partial_\nu \hat{\phi}+\partial_\mu \hat{\sigma}_H \partial_\nu \hat{\sigma}_H\right) \\ &-\frac{\lambda_\chi}{4}\hat{\Omega}^{-4}\hat{\chi}^4-\frac{\lambda_\phi}{4}\hat{\Omega}^{-4}\hat{\phi}^4-\frac{\lambda_H}{4}\hat{\Omega}^{-4}\hat{\sigma}_H^4  \bigg)\,, \label{eq: SI inflation action}
\end{aligned} \end{equation} where the quartic term of the auxiliary scaleron field $ \hat{\chi} $ can be written as \begin{equation}\begin{aligned} 
V=\frac{\lambda_\chi}{4}\Omega^{-4}\hat{\chi}^4=\frac{3}{8}m_P^4 f_0^2 \left[1+\left(1-\frac{\hat{\phi}^2}{6m_P^2}-\frac{\hat{\sigma}_H^2}{6m_P^2}\right)^{-1}\left(\frac{\alpha_\phi}{6m_P^2}\hat{\phi}^2+\frac{\alpha_{H}}{6m_P^2}\hat{\sigma}_H^2\right)\right]\,.
\end{aligned} \end{equation}

\subsection{Inflation parameters}
\label{sec: Inflation parameters}

According to Refs.~\cite{He:2020ivk,Enckell:2018uic,Gundhi:2018wyz}, there is an attractor behaviour, resulting in that the scalar fields (the elementary Higgs scalar and the scaleron) fall into one of the two valleys in the potential of them. This leads to slow roll inflation, driven by the elementary Higgs scalar or the scaleron, corresponding to Higgs or Starobinsky $ R^2 $ inflation, respectively, as the dominant mechanism during inflation. Due to the correlation between the elementary Higgs scalar and the scaleron in the valley, this mixed Higgs/$ R^2 $ inflation can be interpretted as a single field inflation, which can be studied analytically. The scalar fields are in the valley corresponding to Starobinsky inflation driven by the $ R^2 $ term in the region~\cite{He:2020ivk} \begin{equation}\begin{aligned} \label{eq: limits on alphaH R2 inflation}
0\leq \alpha_H\leq\alpha_c/\sqrt{2}\,,
\end{aligned} \end{equation} while in the region\begin{equation}\begin{aligned} 
\alpha_c/\sqrt{2}\leq \alpha_H\leq\alpha_c
\end{aligned} \end{equation} the inflation is in the second valley corresponding to Higgs inflation. Finally, when $ \alpha_H > \alpha_c $, the model is non-perturbative. Here, we define \begin{equation}\begin{aligned} 
\alpha_c \equiv 2.6\times 10^4 \sqrt{\frac{\lambda_H}{0.01}}\,,
\end{aligned} \end{equation} while $ \lambda_H $ is the quartic coupling of $ H $ at the scaleron mass. 

The only constraint on these inflation parameters comes from the normalisation of the scalar amplitude of the primordial curvature fluctuations, which is maintained when~\cite{Gundhi:2018wyz} \begin{equation}\begin{aligned} 
\frac{\lambda_H}{\alpha_H^2+24\lambda_H/f_0^2}\approx 1.2\times 10^{-11}\,. \label{eq: constraint on f0}
\end{aligned} \end{equation} The other constraints from the spectral index of scalar perturbations, $ n_S $, and the tensor-to-scalar ratio, $ r $, are independent of the parameters and both Higgs and $ R^2 $ inflation are in the same universal attractor regime. To leading order in $ 1/N $,~\footnote{$ N $ is the number of e-foldings in Einstein or Jordan frame, since these two frames are the same when the contact terms are added. } these constraints are $ 0.939 < n_S < 0.967 $ and $ 3.8\times 10^{-3}< r < 0.079 $~\cite{Enckell:2018uic}, which are in agreement with data from the Planck collaboration~\cite{Planck:2018jri}. \\

\textbf{Mixed Higgs/$ R^2 $ inflation:} For Higgs inflation, the constraint in Eq.~(\ref{eq: constraint on f0}) requires $ f_0^2 \gg 24\lambda_H/\alpha_H^2$. Due to the fact that higher order corrections to the inflationary parameters diverge when $ \lambda_H/(\alpha_H f_0^2) $ goes against zero~\cite{Enckell:2018uic}, there is an upper bound $ f_0^2 \leq 4\lambda_H /\alpha_H $. Thus, for Higgs inflation, the constraint on $ f_0 $ is given by \begin{equation}\begin{aligned} 
2\times 10^{-5}\ll f_0  \lesssim 3\times 10^{-3} \left(\frac{\lambda_H}{0.01}\right)^{1/4}\,.
\end{aligned} \end{equation} For $ R^2 $ inflation, the condition in Eq.~(\ref{eq: constraint on f0}) requires that $ f_0 \approx 1.7\times 10^{-5} $.
 
After inflation, the scaleron mass is given by $m_\omega= m_P f_0 /\sqrt{2} $. Thus, from the requirement above, this mass is \begin{equation}\begin{aligned} 
3\times 10^{13}~\text{GeV} \ll m_\omega \lesssim  5\times 10^{15}\left(\frac{\lambda_H}{0.01}\right)^{1/4}~\text{GeV}
\end{aligned} \end{equation} for Higgs inflation and $ m_\omega \approx 3\times 10^{13}~\text{GeV} $ for $ R^2 $ inflation.  \\

\textbf{Scale-independent $ R^2 $ inflation:} For scale-invariant inflation, we need that $ f_0\approx 1.7\times 10^{-5} $. In the following, we ignore the VEV of $ \sigma_H $ due to its small value compared to the VEV of $ \phi=v_\phi+\hat{\phi} $, which is given by \begin{equation}\begin{aligned}  \label{eq: phi VEV}
v_\phi=\frac{\sqrt{6}m_P}{\sqrt{1-\alpha_\phi}} 
\end{aligned} \end{equation} for negligible $ \lambda_\phi $. Assuming that the fields $ \phi_N $, $ \sigma_{H,N} $ and $ \sigma_N $ normalize the kinetic terms in Eq.~(\ref{eq: SI inflation action}), then we have that \begin{equation}\begin{aligned} \label{eq: normalized fields}
\phi=-\frac{\alpha_\phi}{1-\alpha_\phi}\phi_N, \quad \quad \sigma_H=\sqrt{-\frac{\alpha_\phi}{1-\alpha_\phi}}\sigma_{H,N}, \quad \quad \sigma=\sqrt{-\frac{\alpha_\phi}{1-\alpha_\phi}}\sigma_N\,.
\end{aligned} \end{equation} Thus, the mass of the normalized field $ \phi_N $ is given by $ m_\phi=\sqrt{1-\alpha_\phi}m_P f_0/\sqrt{2} $. With current constraints on the scalar spectral index, $ n_S $, from the Planck collaboration ($ n_S = 0.9649 \pm 0.0042 $~\cite{Planck:2018vyg,Planck:2018jri}), we obtain the tight bound $ \vert \alpha_\phi \vert < 0.019 $~\cite{Ferreira:2019zzx}. This bound leads to a clear prediction for the tensor-to-scalar ratio $ 0.0026 < r < 0.0033 $~\cite{Ferreira:2019zzx}, which is comfortably within current observational
bounds ($ r \lesssim 0.064 $~\cite{Planck:2018vyg,Planck:2018jri}). This bound on $ \alpha_\phi $ and the fact that $ f_0 \approx 1.7\times 10^{-5} $ result in the mass $ m_\phi \approx 3\times 10^{13}  $~GeV. 

\subsection{The Higgs mass parameter}
\label{sec: The Higgs mass parameter}

In this section, we will determine the magnitude of the the radiative corrections to the mass $ m_H $ of the elementary Higgs $ H $ (introduced in Eq.~(\ref{Eq: fundamentalLagrangian})) for the three inflation model examples. By assuming that non-minimal couplings of the Ricci scalar to the $ \mathbb{Z}_2 $--odd and --even scalar doublet, respectively, are approximately equal (i.e. $ \alpha_{H_\nu}\approx \alpha_{H_\nu} $), the loop-induced mass parameter of $ H_\nu $ will be $ m_{H_\nu}\approx m_H $. For the scale-invariant case, additional mass contributions can also be transferred from the VEV $ v_\phi $, given by Eq.~(\ref{eq: phi VEV}), to the masses $ m_H $ and $ m_{H_\nu} $ by having non-negligible couplings of the forms $ H^\dagger H \phi^2 $ and $ H_\nu^\dagger H_\nu  \phi^2 $, respectively. \\

\textbf{Mixed Higgs/$ R^2 $ inflation:} For mixed Higgs/$ R^2 $ inflation, the one-loop induced mass of $ H $ (derived in Ref.~\cite{Hill:2021psc}) is given by \begin{equation}\begin{aligned} 
\delta m_H^2 = \frac{1}{192\pi^2}m_P^2 f^4_0 \alpha_H(\alpha_H-3)\ln\left(\frac{m_P}{m_\omega}\right)\,. \label{eq: one-loop induced mass of Higgs Higgs R2}
\end{aligned} \end{equation} According to the calculations of these radiative corrections in Ref.~\cite{Hill:2021psc}, the elementary doublet $ H_\nu $ can also achieves such an induced mass without affecting the inflation physics if we add a new non-minimal coupling $ \alpha_{H_\nu} $ of the form $ H_\nu^\dagger H_\nu R $ to the action in Eq.~(\ref{eq: mixed R2/Higgs inflation action}). Thus, the induced mass of $ H_\nu $ will be $ \delta m_{H_\nu} \sim \delta m_H$ for $ \alpha_{H_\nu} \sim \alpha_H $. 

For Higgs inflation, this induced mass of $ H $ is constrained by the various constraints discussed in the previous section. From these constraints, we have $ \delta m_H \gg 1\times 10^{12}\sqrt{\lambda_H/0.01}  $ GeV and $ \delta m_H \lesssim 3\times 10^{16} $~GeV for $ \lambda_H =0.01 $. Moreover, the compositeness scale $ \Lambda_{\rm HC}\approx 4\pi f $ can be pushed up to  the Planck scale with $ m_H =f $ if $ \lambda_H \gtrsim 0.05  $. From the vacuum analysis in Section~\ref{sec: RG analysis and vacuum stability}, the quartic coupling $ \lambda_H $ evaluated at the scaleron mass of $ \mathcal{O}(10^{13}) $~GeV is typical between $ 0.005\dots 0.01 $. Therefore, for the case of Higgs inflation, the induced Higgs mass is constrained to the interval \begin{equation}\begin{aligned} 
1\times 10^{12}\text{ GeV} \ll \delta m_H \lesssim 3\times  10^{16}\text{ GeV}\,. \label{eq: Higgs inflation mH constrained}
\end{aligned} \end{equation} However, in this inflation case, it is necessary that the Higgs vacuum is stable. Therefore, according to upper panel in Figure~\ref{fig: RG running without with neutrinos}, this inflation mechanism can be realized in the TNH model without neutrinos for $ t_\beta \gtrsim 9 $. On the other hand, for the model with neutrinos (see lower panel in Figure~\ref{fig: RG running without with neutrinos}), there is a restricted area in the $ t_\beta $--$ f $ parameter space, where Higgs inflation may be realized. As depicted, there are only few parameter points in the parameter space providing stable vacua and fulfilling the constraints in Eq.~(\ref{eq: Higgs inflation mH constrained}), perhaps for the case with $ m_H=f \sim 10^{14} $~GeV. Therefore, Higgs inflation may be possible to realize in this model framework, but only for a constrained parameter space. 

For $ R^2 $ inflation, the loop-induced mass parameter of $ H $ in Eq.~(\ref{eq: one-loop induced mass of Higgs Higgs R2}) is approximate given by \begin{equation}\begin{aligned} 
\delta m_H \approx (5.5\times 10^7~\text{GeV}) \sqrt{\alpha_H (\alpha_H-3)}\,. \label{eq: R2 inflation mH constrained}
\end{aligned} \end{equation} For this inflation scenario, it is possible that the vacuum is in the second minimum provided one is driven to the unstable, but long-lived minimum at the EW scale. For the central value of the top mass $ 172.13^{+0.76}_{-0.77} $~GeV set by CMS~\cite{CMS:2021jnp}, the Higgs vacuum instability is triggered for $ \alpha_H > 11-12 $~\cite{Figueroa:2017slm}. Futhermore, a lower bound $ \alpha_H > 0.06 $~\cite{Figueroa:2017slm} at the scaleron mass scale is needed to overcome the second minimum during inflation. Therefore, if the vacuum is metastable the loop-induced mass should be $ \delta m_H < (5-6)\times 10^8~\text{GeV} $. According to lower panel in Figure~\ref{fig: RG running without with neutrinos} including neutrinos, few parameter points around $ t_\beta = 5 $ (with $ f = m_H < (5-6)\times 10^8~\text{GeV} $) allow metastable vacua. On the other hand, it is easier to achieve a stable vacuum by requiring $ t_\beta \gtrsim 9 $ and $ m_H\gtrsim 3\times 10^9 $~GeV (obtained by demanding $ \alpha_H \gtrsim 50 $). Finally, without neutrinos, a realization of $ R^2 $ inflation is much less constrained since there are much more parameter points below the required limit $ f < (5-6)\times 10^8~\text{GeV} $ with metastable vacua and thus smaller values of $ t_\beta $. The loop-induced Higgs mass parameter is, however, constrained from above $ \delta m_H \lesssim 1\times 10^{12} $ by the upper limit $ \alpha_H \lesssim \alpha_c/\sqrt{2} $ in Eq.~(\ref{eq: limits on alphaH R2 inflation}) ensuring the $ R^2 $ inflation. \\

\textbf{Scale-independent $ R^2 $ inflation:} In the scale-invariant inflation scenario, a one-loop induced mass parameter of $ H $ is also derived in Ref.~\cite{Hill:2021psc}, which is given by
\begin{equation}\begin{aligned} 
\delta m_H^2 = \frac{1}{192\pi^2}m_P^2 f^4_0 \alpha_H(\alpha_H-\alpha_\phi)(1-\alpha_\phi)\ln\left(\frac{m_P}{m_\phi}\right)\,.
\end{aligned} \end{equation} By using the constraints discussed in the previous section, we obtain that \begin{equation}\begin{aligned} 
\delta m_H \approx (5\times 10^7~\text{GeV})\alpha_H\,, \label{eq: SI R2 inflation mH constrained}
\end{aligned} \end{equation}  which is maintained when $ \alpha_H\gtrsim 1 $ by the constraint $ \vert \alpha_\phi \vert < 0.019 $~\cite{Ferreira:2019zzx}. Furthermore, we have $ \delta m_H \lesssim 1\times 10^{12} $~GeV from the upper limit $ \alpha_H \lesssim \alpha_c/\sqrt{2} $ in Eq.~(\ref{eq: limits on alphaH R2 inflation}). Therefore, these limits lead to the same viable areas in the parameter space in Figure~\ref{fig: RG running without with neutrinos} as in the case of scale dependent $ R^2 $ inflation. However, to achieve scale-invariant $ R^2 $ inflation, we need that all the mass-dependent parameters (also the vector-like masses of the hyperfermions, $ m_{1,2,3} $, and the Majorana masses of the RHNs, $ M_i $) are generated by a scale-invariant UV mechanism.

\begin{figure}[t!]
	\centering
	\includegraphics[width=0.85\textwidth]{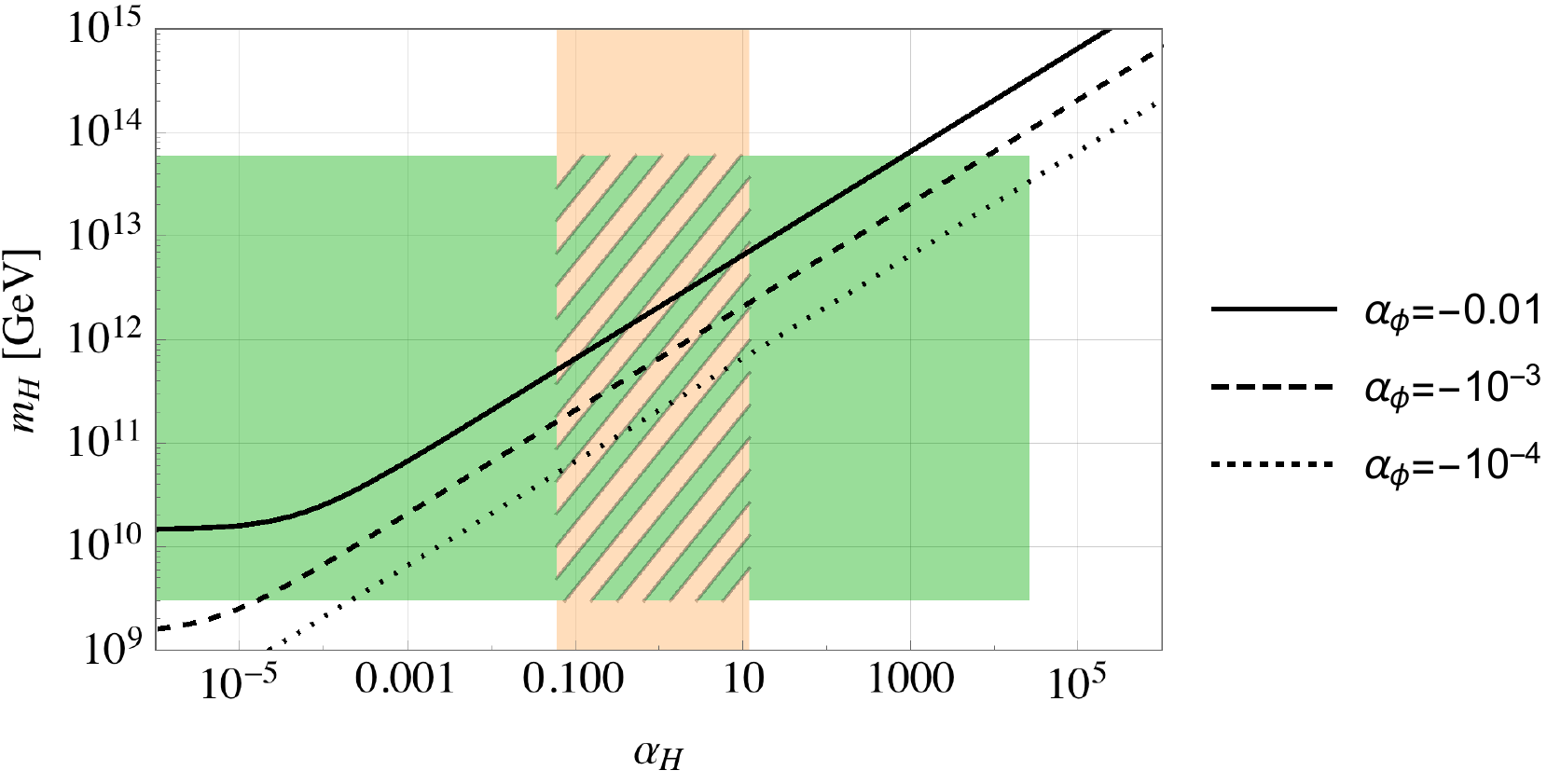} 
	\caption{The mass parameter of the elementary Higgs, $ H $, as function of the non-minimal coupling $ \alpha_H $ for various $ -\alpha_\phi = 0.01,10^{-3},10^{-4} $. According to Figure~\ref{fig: RG running without with neutrinos}, parameter points in the orange-green striped region can both lead to a metastable or stable vacuum, while the orange and green regions require a metastable and stable vacuum, respectively. The unshaded regions give rise to either an unstable vacuum or that one of the neutrino Yukawa couplings is outside the range $ 0.01 \leq h^{ij} \leq 10 $. }
	\label{fig: mH alphaH}
\end{figure}

These mass-dependent parameters can be induced in a simple way via Yukawa couplings of the hyper-fermions and the RHNs to the scalar $ \phi $ of the form $ (\psi_1 \psi_2 + \psi_3 \psi_4+\psi_5 \psi_6)\phi $ and $ N_{R,i} N_{R,i} \phi $, respectively. A smallness of these Yukawa couplings is technically natural due to global chiral symmetries of these fermions. Moreover, by adding a mixing term of the form $ \sigma_H^2 \phi^2 $ to the action in Eq.~(\ref{eq: SI action 1}), a larger mass parameter $ m_H $ can be induced compared to its loop-induced mass, which will open up the parameter space further. In same way, the mass parameter of $ H_\nu $ can be generated by adding a term of the form $  H_\nu^\dagger H_\nu \phi^2 $. This results in a modified version of the action in Eq.~(\ref{eq: SI inflation action}), given by \begin{equation}\begin{aligned} 
S=\int \sqrt{-g}\bigg(&\frac{1}{2}\hat{\Omega}^{-2}g^{\mu\nu}\left(\partial_\mu \hat{\sigma} \partial_\nu \hat{\sigma}+\partial_\mu \hat{\phi} \partial_\nu \hat{\phi}+\partial_\mu \hat{\sigma}_H \partial_\nu \hat{\sigma}_H+\partial_\mu \hat{H}_\nu^\dagger \partial_\nu \hat{H}_\nu\right) \\ &+\frac{1}{2} g^{\mu\nu}\partial_\mu \theta \partial_\nu \theta+\frac{m_P^2}{2} R -\frac{\lambda_\chi}{4}\hat{\Omega}^{-4}\hat{\chi}^4-\frac{\lambda_\phi}{4}\hat{\Omega}^{-4}\hat{\phi}^4-\frac{\lambda_H}{4}\hat{\Omega}^{-4}\hat{\sigma}_H^4  \\ &-\frac{\lambda_{H_\nu}}{4}\hat{\Omega}^{-4}(\hat{H}_\nu^\dagger\hat{H}_\nu)^2-\frac{\lambda_{H\phi}}{4} \hat{\Omega}^{-4} \hat{\sigma}_H^2 \hat{\phi}^2-\frac{\lambda_{H_\nu\phi} }{4} \hat{\Omega}^{-4} \hat{H}_\nu^\dagger \hat{H}_\nu \hat{\phi}^2\bigg)\,, \label{eq: SI inflation action updated}
\end{aligned} \end{equation} where \begin{equation}\begin{aligned} 
\hat{\Omega}^2= 1-\frac{\hat{\phi}^2}{6m_P^2}-\frac{\hat{\sigma}_H^2}{6m_P^2}-\frac{\hat{H}_\nu^\dagger \hat{H}_\nu}{6m_P^2}\equiv \exp\left(\frac{2\theta}{\sqrt{6}m_P}\right)\,,
\end{aligned} \end{equation} and \begin{equation}\begin{aligned} 
\frac{\lambda_\chi}{4}\Omega^{-4}\hat{\chi}^4=\frac{3}{8}m_P^4 f_0^2 \left[1+\hat{\Omega}^{-2}\left(\frac{\alpha_\phi}{6m_P^2}\hat{\phi}^2+\frac{\alpha_{H}}{6m_P^2}\hat{\sigma}_H^2+\frac{\alpha_{H_\nu}}{6m_P^2}\hat{H}_\nu^\dagger \hat{H}_\nu \right)\right]\,.
\end{aligned} \end{equation} 

We assume that the fields $ \phi_N $, $ \sigma_{H,N} $, $ H_{\nu,N} $ and $ \sigma_N $ normalize the kinetic terms in the modified action in Eq.~(\ref{eq: SI inflation action updated}), which are given by \begin{equation}\begin{aligned} \label{eq: normalized fields modified}
&\phi=-\frac{\alpha_\phi}{1-\alpha_\phi}\phi_N, \quad \quad \sigma_H=\sqrt{-\frac{\alpha_\phi}{1-\alpha_\phi}}\sigma_{H,N}, \quad \\ &H_\nu=\sqrt{-\frac{\alpha_\phi}{1-\alpha_\phi}}H_{\nu,N}, \quad\quad \sigma=\sqrt{-\frac{\alpha_\phi}{1-\alpha_\phi}}\sigma_N\,.
\end{aligned} \end{equation} Thus, the above action gives rise to a VEV and mass of $ \phi $, which can be written as \begin{equation}\begin{aligned} 
&v_\phi=\sqrt{\frac{6f_0^2 m_P^2\alpha_\phi}{f_0^2(1-\alpha_\phi)\alpha_\phi-24\lambda_\phi}}\,, \quad \quad \\ & m_\phi^2= \frac{f_0^2 m_P^2 \alpha_\phi^3(f_0^2(1-\alpha_\phi)\alpha_\phi-24\lambda_\phi)^3}{2(1-\alpha_\phi)^2(f_0^2\alpha_\phi^2+24\lambda_\phi)^3} \,, \label{eq: VEV and mass of phi}
\end{aligned} \end{equation} while the mass parameters of $ H $ and $ H_\nu $ are given by \begin{equation}\begin{aligned} 
&m_H^2=m_{H_\nu}^2=\frac{3 f_0^2m_P^2\alpha_\phi(f_0^2(1-\alpha_\phi)\alpha_\phi-24\lambda_\phi)(2\alpha_H\lambda_\phi-\alpha_\phi\lambda_{H\phi})}{(1-\alpha_\phi)(f_0^2\alpha_\phi^2+24\lambda_\phi)^2}\,.
\end{aligned} \end{equation} 

\begin{figure}[t!]
	\centering
	\includegraphics[width=0.95\textwidth]{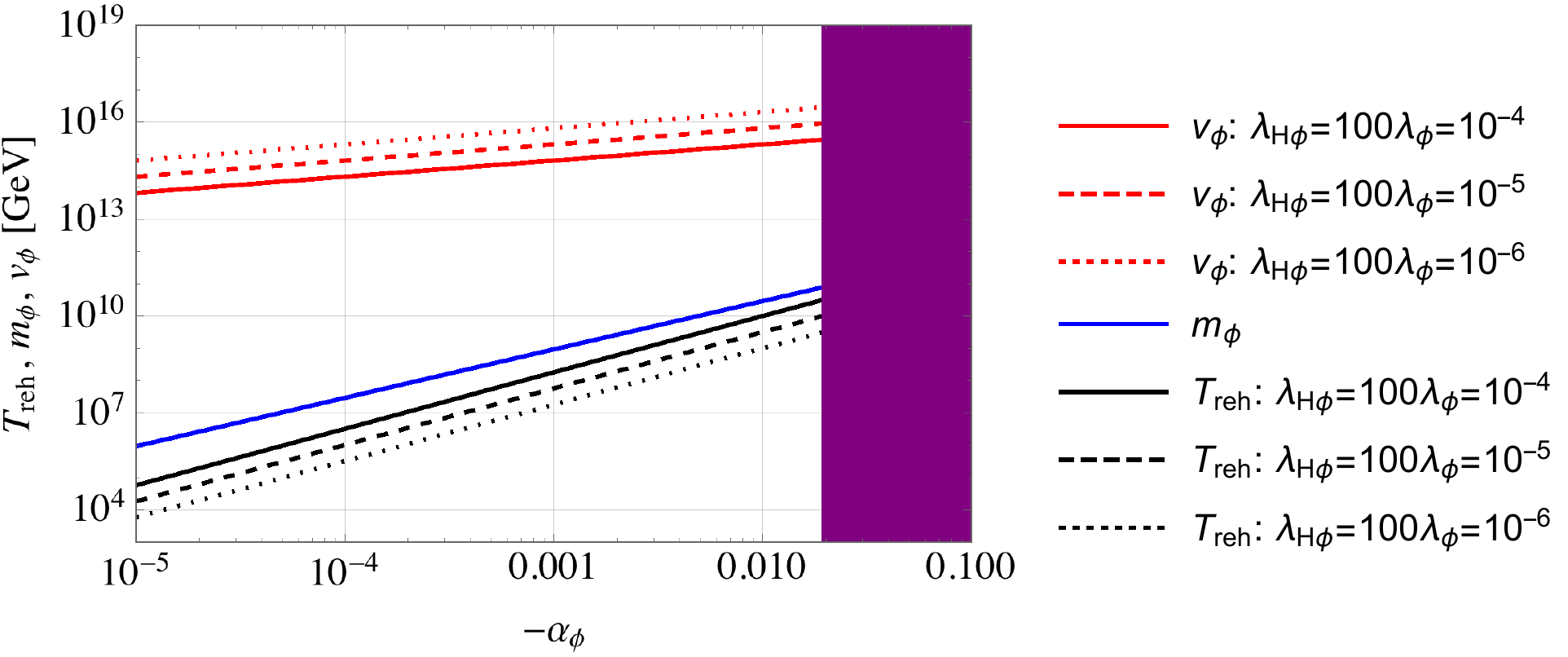} 
	\caption{The VEV of the scaleron $ v_\phi $ (red lines) and the reheating temperature $ T_{\rm reh} $ (black lines) as function of the non-minimal coupling $ \alpha_\phi $ for $  \lambda_{H\phi} = 100\lambda_{\phi}=10^{-4},10^{-5},10^{-6} $, ensuring the slow-roll conditions $ \lambda_{H}\gg \lambda_{H\phi} \gg \lambda_\phi $. The blue line represents the mass of the inflaton, $ m_\phi $. The purple shaded area is excluded by the constraint $ \vert \alpha_\phi \vert < 0.019 $ set by the Planck collaboration given in Ref.~\cite{Planck:2018vyg,Planck:2018jri}.  }
	\label{fig: vphi alphaphi}
\end{figure}

In Figure~\ref{fig: mH alphaH}, the mass parameter $ m_{H} = m_{H_\nu}= f $ is shown for varying $ \alpha_H $ for various $ -\alpha_\phi = 0.01,10^{-3},10^{-4} $ (obeying the constraint $ \vert \alpha_\phi \vert < 0.019 $~\cite{Ferreira:2019zzx}). The parameter points in the orange-green striped area are allowed for both a stable or metastable vacuum, while the points in the orange and green areas require a metastable and stable vacuum, respectively. The unshaded regions lead to either an unstable vacuum or that one of the neutrino Yukawa couplings is outside the range $ 0.01 \leq h^{ij} \leq 10 $. These different regions are determined from the stability analysis gathered in Figure~\ref{fig: RG running without with neutrinos}. Therefore, for $ \alpha_H = \mathcal{O}(1) $, we obtain that the mass parameter $ m_H=\mathcal{O}(10^{12}) $~GeV, where both a stable and metastable vacuum can be obtained. To enforce the slow-roll conditions for this scenario, according to Refs.~\cite{Ferreira:2018qss,Ferreira:2019zzx,Ferreira:2016wem}, we need the hierarchy of the couplings $ \lambda_{H}\gg \lambda_{H\phi} \gg \lambda_\phi $ where $ \lambda_{H} \sim 0.01 $. In the absence of gravity, this hierarchy of these scalar couplings is technically natural due to the underlying shift symmetry of the scale-invariant scalar potential in Eq.~(\ref{eq: SI inflation action updated}). This shift symmetry is broken by the non-minimal couplings, but a calculation of gravitational radiative corrections are required to determine whether the hierarchy survives. We have left the execution of such calculations to future work.

The red lines in Figure~\ref{fig: vphi alphaphi} represent the VEV $ v_\phi $ of $ \phi $ (given in Eq.~(\ref{eq: VEV and mass of phi})) as function of the non-minimal coupling $ \alpha_\phi $ for various scalar couplings $ \lambda_{H\phi}=100\lambda_\phi =10^{-4},10^{-5},10^{-6} $, fulfilling the slow-roll conditions by ensuring the hierarchy $ \lambda_{H}\gg \lambda_{H\phi} \gg \lambda_\phi $. The purple shaded area is excluded by the constraint $ \vert \alpha_\phi \vert < 0.019 $ set by the Planck collaboration~\cite{Planck:2018vyg,Planck:2018jri}. For $ \lambda_{H\phi}=100\lambda_\phi =10^{-4} $ and $\vert\alpha_\phi\vert \sim \mathcal{O}(0.01) $, we have $ v_\phi = \mathcal{O}(10^{15}) $~GeV. Thus, according to Figure~\ref{fig: mH alphaH}, the Yukawa couplings to $ \phi $ providing the RHN masses, $ M_i $, and the vector-like masses, $ m_i $, should be $ \mathcal{O}(10^{-2}) $ for $ \alpha_H =\mathcal{O}(10) $ (leading to $ M_i,m_i \sim \mathcal{O}(10^{13}) $~GeV) and $ \mathcal{O}(10^{-3}) $ for $ \alpha_H =\mathcal{O}(1) $ ($ M_i,m_i \sim \mathcal{O}(10^{12}) $~GeV). The smallness of these Yukawa couplings is technically natural due to global chiral symmetries of these fermions. However, if $ \alpha_H > 11-12 $~\cite{Figueroa:2017slm}, the vacuum has to be stable. According to Figure~\ref{fig: mH alphaH} and~\ref{fig: vphi alphaphi}, in this stable scenario, the Yukawa couplings can take values up to $ \mathcal{O}(0.1) $ with $ M_i,m_i < 10^{14} $~GeV. For smaller $  \lambda_{H\phi}=100\lambda_\phi $ resulting in a larger $ v_\phi $, smaller Yukawa couplings are needed. 

 \subsection{Reheating}
\label{sec: Reheating}

Now, we will consider what happens at the end of inflation for the three inflation scenarios discussed in previous section. At a certain temperature after the inflationary period, the inflaton will begin to decay into lighter particles, resulting in reheating of the Universe. In Ref.~\cite{Aoki:2022dzd}, they found this reheating temperature to be $ \sim 10^{14} $~GeV for Higgs inflation and $ \sim 10^{13} $~GeV for Starobinsky $ R^2 $ inflation, while in the following we will calculate the reheating temperature for the scale-invariant model example including a non-negligible coupling $ \lambda_{H\phi} $. 

At the end of the scale-invariant inflation, the inflaton $ \phi $ will begin to decay dominantly into a pair of either the SM-like Higgs, $ h_1 $  defined in Eq.~(\ref{Eq: Higgs states}), or the dilaton, $ \sigma $, where the decay rates into fermions are suppressed by the factor $ (m_f/m_P)^2 $~\cite{Gorbunov:2010bn}. These processes lead to the reheating of the Universe after the inflation in which the fields acquire large kinetic energy which is subsequently rapidly dampened by expansion. Here we define the reheating temperature, $ T_{\rm reh} $, as the temperature the radiation would have at the time when the Universe becomes radiation dominated if it were in thermal equilibrium.

The reheating temperature is given by \begin{equation}\begin{aligned} 
T_{\rm reh}=\left(\frac{90}{g_*\pi^2}\right)^{1/4}\sqrt{m_P \Gamma_{\rm tot}}\,, \label{eq: reheating temperature}
\end{aligned} \end{equation} where $  \Gamma_{\rm tot} $ is the total inflaton decay rate and the number of relativistic species at the inflaton decay is given by the SM value, $ g_*=106.75 $. To tree-level, the dominant decay rate of $ \phi $ is into a pair of the observed SM-Higgs $ h_1 $ in Eq.~(\ref{Eq: Higgs states}), which is given by \begin{equation}\begin{aligned} 
\Gamma_{\phi h_1 h_1}=\frac{g_{\phi h_1h_1}^2}{8\pi m_\phi}
\end{aligned} \end{equation} with the decay coupling of $ \phi $ into a pair of $ h_1 $ \begin{equation}\begin{aligned} 
g_{\phi h_1h_1}&=c_\alpha^2\frac{\alpha_\phi^2\left[f_0^2(1-\alpha_\phi)\alpha_\phi-24\lambda_\phi\right]^2}{4\sqrt{6}(1-\alpha_\phi)^2(f_0^2\alpha_\phi^2+24\lambda_\phi)^3}\sqrt{\frac{f_0^2m_P^2\alpha_\phi}{f_0^2(1-\alpha_\phi)\alpha_\phi-24\lambda_\phi}} \bigg(288\lambda_{H\phi}\lambda_\phi+\\ & f_0^4(1-\alpha_H)(1-\alpha_\phi)\alpha_\phi^2+12f_0^2\Big[(\alpha_\phi-2)\alpha_\phi\lambda_{H\phi}+2(\alpha_H-(1-\alpha_H)\alpha_\phi)\lambda_\phi\Big]\bigg)\,.
\end{aligned} \end{equation} Here $ \alpha $ is the mixing angle between the elementary and composite Higgs states in Eq.~(\ref{Eq: Higgs states}), where $ c_\alpha \approx 1 $. Finally, the coupling of $ \phi $ into dilatons is $ g_{\phi \sigma\sigma}\approx g_{\phi h_1 h_1}/2 $, and thus the inflaton decay rate associated with these decays is $ \Gamma_{\phi \sigma\sigma}\approx \Gamma_{\phi h_1 h_1}/4 $.

In Figure~\ref{fig: vphi alphaphi}, the inflaton mass $ m_\phi $ (the blue line) is compared with the reheating temperature $ T_{\rm reh} $ (the black lines), given by Eq.~(\ref{eq: reheating temperature}), as function of non-minimal coupling $ \alpha_\phi $ for various couplings $ \lambda_{H\phi} = 100\lambda_{\phi}=10^{-4},10^{-5},10^{-6} $, where the purple shaded area is excluded by the Planck collaboration~\cite{Planck:2018vyg,Planck:2018jri}. For the entire interesting region studied in the previous section, the reheating temperature is always smaller than the inflaton mass. For $ \lambda_{H\phi}=100\lambda_\phi =10^{-4} $ and $\vert\alpha_\phi\vert \sim \mathcal{O}(0.01) $, the reheating temperature is $ T_{\rm reh}\sim \mathcal{O}(10^{10}) $~GeV, while the inflaton mass is $ m_\phi \sim 3T_{\rm reh}  $. \\

In this section, we have addressed the three inflation scenarios under consideration and demonstrate their ability to resolve the naturalness problem outlined in Ref.~\cite{Ferreira:2021ctx}. This problem arises from the significant hierarchy between a large loop-induced mass parameter of the Higgs boson, $ m_H $, and its observed value. Importantly, we show that the TNH framework offers a viable solution to this naturalness issue, even when accounting for neutrino masses. In the subsequent section, we will delve into the quest to elucidate the observed matter-antimatter asymmetry in the Universe through thermal leptogenesis. We will explore how the scotogenic neutrino sector, integrated into these specific TNH models, may provide valuable insights into this phenomenon.

 \section{Leptogenesis}
\label{sec: Leptogenesis}

The baryon asymmetry of the Universe (BAU) cannot be explained within the SM of particle physics. The amount of BAU is typical quantified by the cosmic baryon-to-photon ratio $ \eta_B^{\rm obs}\simeq 6.1\times 10^{-10} $~\cite{ParticleDataGroup:2016lqr,Planck:2015fie}. An appealing way to dynamically generate the BAU is baryogenesis via leptogenesis in the early Universe~\cite{Fukugita:1986hr}. In this section, we show that the scotogenic neutrino sector of this model  may provide thermal leptogenesis due to the fact that a CP asymmetry is generated by RHN decays, $ N_i \rightarrow l_{L,i} H_\nu, \overline{l}_{L,i} H_\nu^* $. Finally, the EW sphalerons transfer this lepton asymmetry into a baryon asymmetry. There exist various studies of thermal leptogenesis in the
scotogenic model in the literature. It was first time proposed by E.~Ma in Ref.~\cite{Ma:2006fn} and studied in more detail subsequently in Refs.~\cite{Kashiwase:2012xd,Kashiwase:2013uy,Racker:2013lua,Clarke:2015hta,Hugle:2018qbw}. According to Ref.~\cite{Hugle:2018qbw}, none of the leptogenesis studies in Refs.~\cite{Kashiwase:2012xd,Kashiwase:2013uy,Racker:2013lua,Clarke:2015hta} seem completely exhaustive. Therefore, in the following, we will focus on the results of the leptogenesis analysis done in Ref.~\cite{Hugle:2018qbw}. A more comprehensive summary of leptogenesis studies in the scotogenic model can be found in Ref.~\cite{Cai:2017jrq}.  

The observed value of BAU can be produced, if the lightest RHN mass is given by~\cite{Hugle:2018qbw}\begin{equation}\begin{aligned} 
M_1\approx \frac{\xi_1}{\xi_3}\frac{m_l}{m_h}2\times 10^{11}~\text{GeV}\,,\label{eq: M1 leptogenesis}
\end{aligned} \end{equation} where $ m_{l,h} $ are the masses of the lightest and heaviest active neutrino, respectively, and we define \begin{equation}\begin{aligned} 
\xi_i \equiv \left[\frac{1}{8}\frac{M_i^2}{m_R^2-m_I^2}\left(\frac{m_R^2}{m_R^2-M_i^2}\ln \left(\frac{m_R^2}{M_i^2}\right)-\frac{m_I^2}{m_I^2-M_i^2}\ln \left(\frac{m_I^2}{M_i^2}\right)\right)\right]^{-1}\,.
\label{eq: xi i}
\end{aligned} \end{equation} Here $ m_{R,I} $ are the masses of the mass eigenstates $ \widetilde{\sigma}_{R,I} $ of the mass matrices in Eq.~(\ref{Eq: MR MI mass matrices}), consisting mostly of the neutral components $ \sigma_{R,I} $ in Eq.~(\ref{Eq: elementary doublet neutrino}), respectively. This analytical expression is valid for $ m_l \lesssim 10^{-3} $~eV down to $ m_l \approx 10^{-6} $~eV where $ \Delta L =2 $ washout becomes relevant. Below this point ($ m_l \lesssim 10^{-6} $~eV), according to Ref.~\cite{Hugle:2018qbw}, the mass $ M_1 $ must be larger compared to the approximation given by Eq.~(\ref{eq: M1 leptogenesis}) -- but not larger than for $ m_l > 10^{-6} $~eV -- to produced the observed BAU. In the following, we consider the production of BAU for the three inflation scenarios: \\

\begin{figure}[t!]
	\centering
	\includegraphics[width=0.95\textwidth]{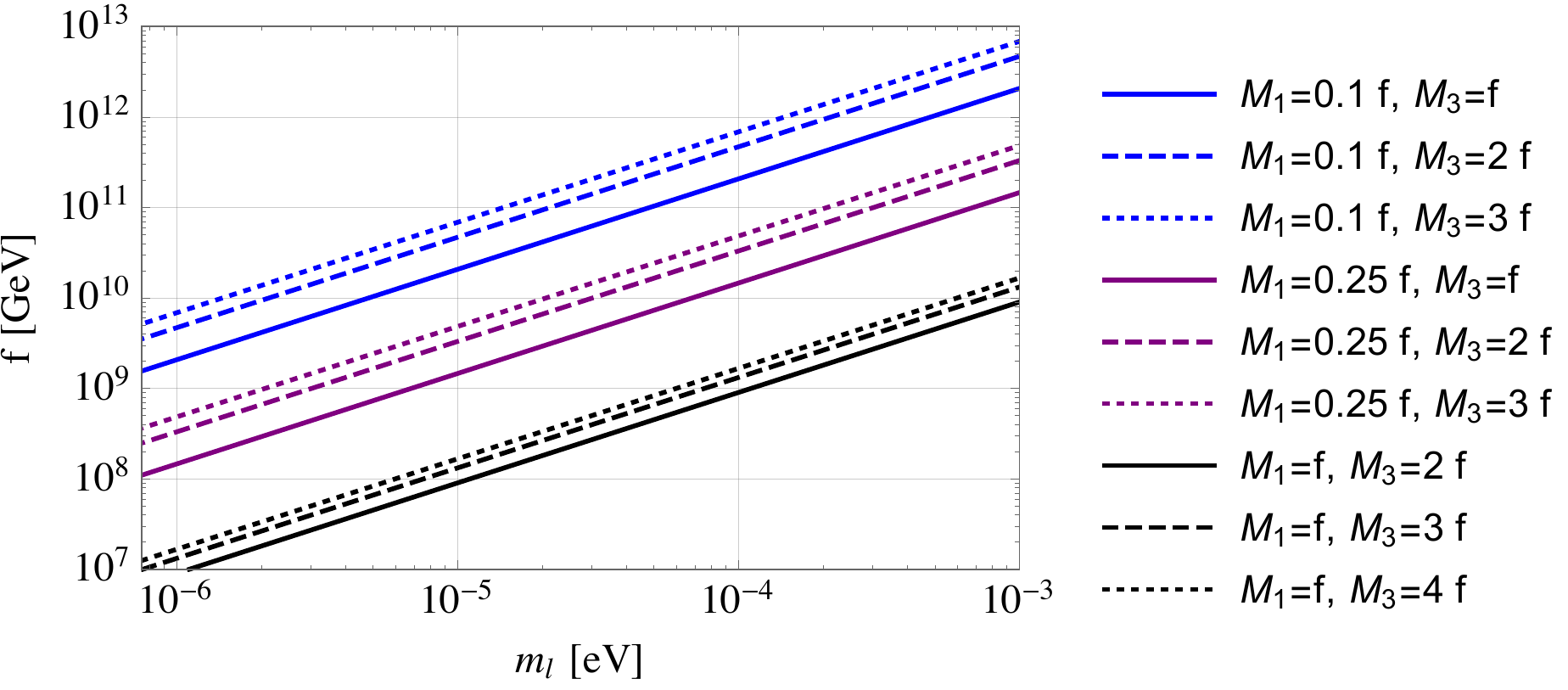} 
	\caption{The decay constant $ f $ producing the observed value $ \eta_B^{\rm obs}\simeq 6.1\times 10^{-10}  $~\cite{ParticleDataGroup:2016lqr,Planck:2015fie} as function of the lightest neutrino mass $ m_l $ for various RHN masses $ M_{1,3} $, where we assume the hierarchy $ M_1 < M_{2} \lesssim M_3 $.}
	\label{fig: f vs ml leptogenesis}
\end{figure}

\textbf{Mixed Higgs/$ R^2 $ inflation:} In Figure~\ref{fig: f vs ml leptogenesis}, the decay constant $ f $ producing the observed value $ \eta_B^{\rm obs}\simeq 6.1\times 10^{-10}  $~\cite{ParticleDataGroup:2016lqr,Planck:2015fie} is plotted as function of the lightest neutrino mass $ m_l $ for various RHN masses $ M_{1,3} $, where we assume the hierarchy $ M_1 < M_{2} \lesssim M_3 $. For Higgs inflation, it is difficult to produce small enough asymmetry due to the fact that we need $ f = m_H \gg 10^{12} $~GeV as found in Eq.~(\ref{eq: Higgs inflation mH constrained}) and, therefore, we need $ M_1 \ll f $ giving rise to small neutrino Yukawa couplings $ h^{ij}\ll 0.01 $. However, for the $ R^2 $ inflation case, it can be realized both for $ f <(5-6)\times 10^8 $~GeV ($ t_\beta \approx 5 $ and $ \alpha_H=\mathcal{O}(1) $) allowing metastable vacuum and $ f \gtrsim 3\times 10^9 $~GeV requiring a stable vacuum ($ t_\beta \gtrsim 9 $ and $ \alpha_H \gtrsim 50 $). Thus, we have $ M_1=\mathcal{O}(f) $ and $ h^{ij}>0.01 $ and, therefore, we do not need to tune them down to smaller values as in the Higgs inflation case. \\ 

\textbf{Scale-independent $ R^2 $ inflation:} Finally, in the case with scale-invariant inflation, the parameters $ M_{2,3} \approx f =\mathcal{O}(10^{12}) $~GeV, $ M_1 = \mathcal{O}(0.1)f $ and $ m_l \sim 10^{-3} $~eV can provide the observed asymmetry (see the blue solid line in Figure~\ref{fig: f vs ml leptogenesis}). For this example, we have $ \alpha_H =\mathcal{O}(1) $ with $ \alpha_\phi \sim -0.01 $ (according to Figure~\ref{fig: mH alphaH}), while some Yukawa couplings should be $ \mathcal{O}(10^{-3}) $ to generate the masses $ M_{2,3},m_{1,2,3} $ and $ \mathcal{O}(10^{-4}) $ to induce $ M_{1} $ (by considering Figure~\ref{fig: vphi alphaphi}). According to Figure~\ref{fig: mH alphaH}, the vacuum of this example is allowed to be either metastable ($ 2 \lesssim t_\beta \lesssim  16 $) or stable ($ t_\beta \gtrsim 9 $) depending on the value of $ t_\beta $ and the remaining parameters, illustrated by the lower panel in Figure~\ref{fig: RG running without with neutrinos}. In Figure~\ref{fig: RG running examples}, the solid curves show the RG running of $ \lambda_H $ for a metastable (with $ t_\beta=5 $) and stable vacuum ($ t_\beta=20 $), where $ f=10^{12} $~GeV and $ M_{2,3}=10M_1=f $ as in this example. Such a model with these parameters may explain the BAU. \\

In the next section, we will investigate the possibility for a composite super-heavy DM candidate in these concrete TNH models, non-thermally produced by the weakness of the gravitational interactions.

 \section{Composite super-heavy dark matter}
\label{sec: Dark matter}

Now, we consider the possibility that the DM may be composite super-heavy DM (SHDM) candidates in this model framework, heavier than the inflaton, where its relic density today is non-thermally produced by the weakness of the gravitational interactions~\cite{Chung:1998zb,Kuzmin:1998uv,Kuzmin:1998kk,Kolb:1998ki,Chung:1999ve,Kuzmin:1999zk,Chung:2001cb}. As long as there are stable particles with mass of the order
of the inflaton mass, the sufficient abundance will be generated quite independently of the non-gravitational interactions of the DM candidates. Indeed, this non-thermal gravitational production of the DM abundance during inflation is the only experimentally verified DM production mechanism, since the observed CMB fluctuations have exactly the same origin~\cite{Kannike:2016jfs}. Therefore, it seems that SHDM is the most natural candidate of DM in this model framework due to the fact that the other DM candidates have no experimental verification yet. 

This may be realized simply by extending the spectrum of hyper-fermions in Table~\ref{tab:fermionssu6sp6} with only one extra Dirac (two Weyl) fermion, called $ \lambda $, transforming adjoint under the new strong gauge group $ G_{\rm HC}=\SU(2)_{\rm HC} $, which is the real representation. We can arrange these two Weyl spinors in a vector transforming under the fundamental representation of the $ \SU(2)_\Lambda $ chiral symmetry group: $ \Lambda=(\lambda_L,\widetilde{\lambda}_L)^T $ with $ \widetilde{\lambda}_L \equiv \epsilon \lambda_R^* $. Altogether, the hyper-fermion sector features the global (chiral) symmetry $ \SU(6)_\Psi\otimes \SU(2)_\Lambda\otimes \UU(1)_\Theta $ at the quantum level. We add to the underlying Lagrangian in Eq.~(\ref{Eq: fundamentalLagrangian}) an underlying fermionic Lagrangian of $ \Lambda $, which can be written as \begin{equation}\begin{aligned}  \label{eq: underlying Lag Lambda-sector}
\mathcal{L}_\Lambda = \overline{\Lambda} i \gamma^\mu \partial_\mu \Lambda - \left(\frac{1}{2}\Lambda^T M_\Lambda\Lambda +\rm h.c.\right)\,,
\end{aligned} \end{equation} where the vector-like mass of $ \Lambda $ are given by the matrix  \begin{equation}\begin{aligned} \label{Eq: vector-like masses}
M_\Lambda &= \begin{pmatrix} 0 &  m_\Lambda \\   m_\Lambda  & 0 \end{pmatrix}\,.
\end{aligned} \end{equation} In the scale-invariant TNH model, this mass can be induced via Yukawa coupling of the form $ y_\lambda \phi \lambda_L \widetilde{\lambda}_L + \rm h.c. $ when the $ \phi $ achieves a VEV, $ v_\phi $, by the spontaneous symmetry breaking of the Weyl symmetry. 

Upon the condensation, the dynamical condensate in Eq.~(\ref{eq: condensate Phi sector}) formed by the hyper-fermions in the $ \Phi $-sector is thus extended by the condensate by the hyper-fermion in the $ \Lambda $-sector, which is given by \beq
\langle \Lambda^A_{\alpha,i}\Lambda^B_{\beta,j}\rangle\epsilon^{\alpha\beta}\delta^{ij}\sim f_\Lambda^3 E^{AB}_{\Lambda}\,, \eeq with $ A,B $ flavour, $ \alpha,\beta $ spin and $ i,j $ HC gauge indices, while the $ \Lambda $-sector vacuum matrix is $ E_\Lambda=\sigma^1 $. We expect the Goldstone boson decay constants $ f $ and $ f_\Lambda $ to be of the same order~\cite{Ryttov:2008xe,Frandsen:2011kt} and will for simplicity take them to be identical. These condensates break these global symmetries $ \SU(6)_\Psi$ to $ \SP(6)_\Psi $ resulting in the composite pNGBs listed in Table~\ref{tab:su6sp6} and $ \SU(2)_\Lambda\otimes \UU(1)_\Theta $ to $ \UU(1)_\Lambda\otimes \mathbb{Z}_2  $ providing one complex and one real composite pNGB, denoted $ \Phi $ and $ \Theta $, respectively. The pNGB matrix $ \Sigma $ given in Eq.~(\ref{eq: GB matrix}) is modified as follows \begin{equation}\begin{aligned}
\frac{\pi^a X^a}{f} \rightarrow \frac{\pi^a X^a}{f}+\frac{1}{3}\frac{\Theta}{f_\Theta}\mathbbm{1}_4\,,
\end{aligned} \end{equation} while the Goldstone excitations from the $ \Lambda $-sector around the vacuum $ E_\Lambda $ is then parameterized by \begin{equation}\begin{aligned}
\Sigma_\Lambda =\text{exp}\left[2\sqrt{2}i\left(\frac{\pi^a_\Lambda X^i_\Lambda}{f}+\frac{1}{6}\frac{\Theta}{f_\Theta}\mathbbm{1}_2\right)\right]E_\Lambda
\end{aligned} \end{equation} with the broken generators $  X^i_\Lambda = \sigma^i  $, $ i=1,2 $. Furthermore, we have the unbroken generator given by the third Pauli matrix $ \sigma^3 $, which is associated with the unbroken $ \UU(1)_\Lambda $ global symmetry. For simplicity, we will, henceforth, use the notations $ \Phi\equiv (\pi^1_\Lambda-i\pi^2_\Lambda)/\sqrt{2} $ and $ \overline{\Phi}\equiv (\pi^1_\Lambda+i\pi^2_\Lambda)/\sqrt{2} $, which have the $ \UU(1)_\Lambda $ charges $ \pm 1 $, respectively. 

Below the condensation scale, the Lagrangian of Eq.~(\ref{eq: underlying Lag Lambda-sector}) yields\begin{equation}\begin{aligned} \label{eq: SHDM eff lag}
\mathcal{L}_{\rm eff,\Lambda}=\frac{f_\Lambda^2}{8}\text{Tr}[\partial_\mu \Sigma_\Lambda^\dagger \partial^\mu \Sigma_\Lambda]-2\pi Z_\Lambda f_\Lambda^3 \text{Tr}[M_\Lambda \Sigma_\Lambda^\dagger +\Sigma_\Lambda M_\Lambda^\dagger]\,,
\end{aligned} \end{equation} where the first term contains the kinetic terms of the pNGBs, while the second term is the effective potential contribution from the $ \Lambda $-sector. Here $ Z_\Lambda $ is a non-perturbative $ \mathcal{O}(1) $ constant that can be suggested by lattice simulations. Among the composite pNGBs, DM may be identified as the composite complex pNGB $ \Phi $, studied in Ref.~\cite{Dietrich:2006cm,Luty:2008vs,Ryttov:2008xe,Frandsen:2011kt,Alanne:2018xli,Rosenlyst:2021elz}, which remains protected from decaying by the unbroken $ \UU(1)_\Lambda $ global symmetry and has the mass \begin{equation}\begin{aligned} \label{Eq: DM mass expression}
m_\Phi^2	=8\pi Z_\Lambda f_\Lambda m_\Lambda\,.
\end{aligned} \end{equation} 

For $ m_\Lambda $ close to $ f $ and $ f_\Lambda \approx f $, the mass of this composite DM candidate is of the order of $ f $. Such a DM candidate may identified as SHDM. From Ref.~\cite{Aloisio:2015lva}, a simple estimate of the ratio of the SHDM density to the critical
density today based on gravitational production is given by \begin{equation}\begin{aligned} \label{eq: DM relic density}
\Omega_{\Phi}(t_0)\equiv \frac{\rho_\Phi(t_0)}{\rho_c(t_0)}\simeq 2\times 10^{-3}\Omega_R \frac{8\pi}{3}\left(\frac{T_{\rm reh}}{T_0}\right)\left(\frac{m_{\chi/\phi}}{m_P}\right)^2\left(\frac{m_\Phi}{m_{\chi/\phi}}\right)^{5/2}\exp\left(-\frac{2m_\Phi}{m_{\chi/\phi}}\right)\,,
\end{aligned} \end{equation} where $ T_0 =2.3\times 10^{-13} $~GeV is the CMB temperature today and $ \Omega_R=4\times 10^{-5} $ is the radiation density today. From Planck observations of the CMB, the mean DM density in the Universe in units of the critical density has been measured to be $ \Omega_{\rm DM}h^2=0.1198\pm 0.0015 $~\cite{Planck:2018vyg}, with $ h=0.678\pm 0.009 ~\text{km}\phantom{.}\text{s}^{-1}\text{Mpc}^{-1} $, resulting in $ \Omega_{\rm DM}=0.261 $. According to Section~\ref{sec: Reheating}, the reheating temperature is $  T_{\rm reh} \sim 10^{14} $~GeV for Higgs inflation and $  T_{\rm reh} \sim 10^{13} $~GeV for Starobinsky $ R^2 $ inflation, while $ T_{\rm reh} $ is given in Figure~\ref{fig: vphi alphaphi} for the scale-invariant inflation scenario as a function of various parameters. \\

\textbf{Mixed Higgs/$ R^2 $ inflation:} If we assume that the vector-like mass $ m_\Lambda = m_{1,2,3} \approx 0.3 f $ and the decay constant $ f_\Lambda \approx f $ in the $ \Lambda $-sector, the DM mass is about $ 4 f $. For the Higgs inflation, we need $ 1\times 10^{13}~\text{GeV}\ll m_\Phi \lesssim  7\times 10^{17}~\text{GeV} $ to obtain the observed DM abundance, which requires $ 3 \times 10^{12}~\text{GeV}\ll f \lesssim 2 \times 10^{17}~\text{GeV} $. This range is further constrained from above by $  f \lesssim 3 \times 10^{16}~\text{GeV} $ in Eq.~(\ref{eq: Higgs inflation mH constrained}), resulting in the DM mass of $ 1\times 10^{13}~\text{GeV}\ll m_\Phi \lesssim  1\times 10^{17}~\text{GeV} $. According to lower panel in Figure~\ref{fig: RG running without with neutrinos}, when scotogenic neutrinos are included for this inflation scenario, we need $ f\sim 10^{14} $~GeV resulting in $ m_\Phi\sim 4\times 10^{14}  $. For the $ R^2 $ inflation, the DM mass may be $ m_\Phi \sim 9\times 10^{9} $~GeV to obtain the observed DM relic density, which requires $ f \sim 2\times 10^{9} $~GeV. According to lower panel in Figure~\ref{fig: RG running without with neutrinos}, this value of $ f\sim 2\times 10^9 $~GeV is in between the two viable regions, resulting in a metastable or stable vacuum, but it can easily be smaller or larger depending on more precise calculations of the reheating temperature, the value of the non-perturbative coefficient $ Z_\Lambda $ in Eq.~(\ref{eq: SHDM eff lag}) and the hierarchy of the vector-like hyper-fermion masses. \\ 

\textbf{Scale-independent $ R^2 $ inflation:} By following the scale-invariant inflation example with $ f=\mathcal{O}(10^{12}) $~GeV (for $ \alpha_H = \mathcal{O}(1) $ and $ \alpha_\phi \sim -0.01 $) alleviating the various problems discussed so far, the scaleron (identified as the inflaton) has the mass $ m_\phi \sim 3\times 10^{10} $~GeV while the reheating temperature is $ T_{\rm reh}\sim 1\times 10^{10} $~GeV according to Figure~\ref{fig: vphi alphaphi}. Thus, the DM mass is $ \mathcal{O}(10^{12}) $~GeV, giving $ m_\Phi/m_\phi \sim 100 $. However, this results in a vanishingly small DM relic density in Eq.~(\ref{eq: DM relic density}) due to the exponential suppression from the large mass ratio $ m_\Phi/m_\phi $. To obtain the observed DM relic abundance from this SHDM candidate ($ \Omega_\Phi(t_0) = \Omega_{\rm DM} $), $ m_\Phi/m_\phi \sim 5 $ is needed. It can be achieved by assuming $ m_\Lambda \sim 10^{-3}f $, but it requires an Yukawa coupling between $ \phi $ and the $ \Lambda $--hyper-fermion of $ y_\Lambda \sim 10^{-6} $. This small Yukawa coupling is technically natural according to G.~'t Hooft's naturalness principle~\cite{tHooft:1979rat} as it reveals the restoration of the global chiral $ \SU(2)_\Lambda $ symmetry when $ y_\Lambda \rightarrow 0 $. The total DM abundance can also be produced for $ m_\Lambda=m_{1,2,3} $ if the non-perturbative constant in Eq.~(\ref{Eq: DM mass expression}) is $ Z_\Lambda \sim 0.01 $. Moreover, this scenario can be tested by lattice calculations of $ Z_\Lambda $. Finally, the observed DM relic density can be obtained by decreasing $ f $ to $ 2\times 10^{11} $~GeV, requiring $ \alpha_H \sim 0.01 $ and a stable vacuum (i.e. $ t_\beta \gtrsim 9 $) according to Figure~\ref{fig: mH alphaH}. This scenario with $f \sim 2\times 10^{11} $~GeV can also alleviate the other problems discussed so far. \\

In conclusion, the observed DM relic abundance in our Universe may be non-thermally produced as SHDM by the weakness of the gravitational interactions in the TNH framework for the three inflation scenarios. For these cases, the DM is identified as a composite complex scalar, $ \Phi $, consisting of only one extra Dirac hyper-fermion, which is protected from decaying by a global $ \UU(1)_\Lambda $ symmetry. If we assume $ m_\Lambda = m_{1,2,3} $, the DM mass is $ 1\times 10^{13}~\text{GeV}\ll m_\Phi \lesssim  1\times 10^{17}~\text{GeV} $ for Higgs inflation, which is consistent with constraints on $ f $. For $ R^2 $ inflation, $ m_\Phi \sim 9\times 10^{9} $~GeV for $ m_\Lambda = m_{1,2,3} $, which may be consistent with constraints on $ f $ depending on more precise calculations of the reheating temperature, the value of the non-perturbative coefficient $ Z_\Lambda $ in Eq.~(\ref{eq: SHDM eff lag}) and the hierarchy of the vector-like hyper-fermion masses. In the scale-invariant inflation scenario, the DM mass is about $ \mathcal{O}(10^{11}) $~GeV. This can be achieved by a technically natural small Yukawa coupling between scaleron and the new hyper-fermion of $ y_\Lambda \sim 10^{-6} $, a non-perturbative constant $ Z_\Lambda \sim 0.01 $, a reduction of $ f $ to $ 2\times 10^{11} $~GeV requiring $ \alpha_H\sim 0.01 $ or a combination of these possibilities.

 \section{QCD axion dark matter}
\label{sec: QCD axion dark matter}

In this section, we will only focus on the scale-invariant inflation scenario, where we will investigate the advantage of assuming that the inflaton, $ \phi $, is a complex scalar. By making this assumption, the strong CP problem~\cite{Peccei:1977hh,Weinberg:1977ma,Wilczek:1977pj} can be solved, while DM may be identified as a so-called QCD axion.~\footnote{However, there are also recently other resolutions~\cite{Ai:2020ptm,Nakamura:2021meh,Yamanaka:2022bfj} of the strong CP problem proposed within QCD and thus we do not need new particles beyond the SM such as a QCD axion and a new QCD coloured quark triplet.} In this case, the inflaton is charged under a global lepton $ \UU(1) $ symmetry, which is spontaneously broken by the VEV of $ \phi $ and its angular part $ a $ will be identified as a NGB. To solve the CP problem, we just have to introduce a new chiral pair of $ \SU(3)_{\rm C} $ coloured triplets $ Q_{L,R} $ with lepton numbers such that they couple to $ \phi $ via Yukawa interactions. This leads to that this lepton $ \UU(1) $ symmetry becomes colour-anomalous, and therefore it is identified as a Peccei-Quinn symmetry and $ a $ becomes the QCD axion. This PQ mechanism~\cite{Peccei:1977hh} thus solves the strong CP problem~\cite{Kim:1979if,Shifman:1979if}, providing a tiny small mass to the axion given by~\cite{Weinberg:1977ma,Wilczek:1977pj,GrillidiCortona:2015jxo} \begin{equation}\begin{aligned} \label{eq: axion mass}
m_a \simeq 0.57\times \left(\frac{10^{13}~\text{GeV}}{f_a}\right)~\mu\text{eV} \,.
\end{aligned} \end{equation} Here the axion decay constant $ f_a $ is identified with the VEV of $ \phi $, i.e. $ f_a = v_\phi $. As we will see, if we have a standard misalignment in the interval of $ 0.01-0.1 $, the axion relic abundance can explain the observed DM relic density.  

Note that if there is more than one extra Dirac quark $ Q $ that contributes to the colour anomaly of the PQ symmetry~\cite{Sikivie:1982qv}, this solution of the strong CP problem leads to a Universe dominated by the energy of very energetic domain walls. Therefore, we only add one extra quark. This extra quark $ Q $ is thus cosmologically stable if it is an EW singlet without hypercharges~\cite{Nardi:1990ku}, resulting in a substantial relic abundance of them hadronises with the SM-quarks and becomes heavy hadrons with electric charge. This is strongly constrained by searches of charged massive stable particles~\cite{Perl:2001xi,Perl:2004qc,Chuzhoy:2008zy,Perl:2009zz}. However, this can be avoided by assuming that $ Q $'s have hypercharges $ -1/3 $ and $ +2/3 $ such that they mix with the SM-quark singlets, which leads to possible co-annihilation and decay of the $ Q $'s. 

Charge assignments of the various fields in this scale-invariant TNH model under the global PQ and lepton $ \UU(1) $ symmetry are $ -1/2 $ for the fermions $ q_L $, $ u_R $, $ d_R $, $ l_L $, $ N_R $, $ e_R $, $ Q $ and $ \widetilde{Q} $, and $ +1 $ for $ \phi $. The remaining fields are not charged under this new symmetry. This new symmetry is thus preserved in the Yukawa interaction terms in Eqs.~(\ref{Eq: Higgs-Yukawa couplings}) and~(\ref{Eq: Higgs-Yukawa couplings with neutrinos}). With these charge assignments, we can also write the Yukawa couplings: \begin{equation}\begin{aligned} 
\mathcal{L}_Y\supset - y_Q \widetilde{Q}\phi Q - y_{Qd}^i Q \phi d_{R,i} +\text{h.c.} \,.
\end{aligned} \end{equation} Notice $ Q $ and $ \widetilde{Q} $ transform, respectively, under the fundamental and antifundamental representation of $ \SU(3)_{\rm C} $ and have the hypercharges $ \mp 1/3 $, where $ Q_{L,R} $ form a Dirac fermion with mass $ m_Q = y_Q v_\phi /\sqrt{2} $. 

For the concrete scale-invariant TNH model with $ f=\mathcal{O}(10^{12}) $~GeV and $ \alpha_\phi \sim -0.01 $, the axion decay constant is $ f_a= v_\phi \sim 10^{15} $~GeV, see Figure~\ref{fig: vphi alphaphi}. According to Figure~1 in Ref.~\cite{Eroncel:2022vjg}, if this Kim-Shifman-Vainshtein-Zakharov-like (KSVZ-like) axion discussed above is produced from the standard misalignment mechanism with an initial misalignment angle $ a_i =\mathcal{O}(1) $, it leads to an overproduction of DM. However, in the case where $ a_i $ between $ 0.01 $ and $ 0.1 $, this axion  accounts for the cold DM in the Universe and corresponds to a QCD axion with the mass $ m_a \sim 6\times 10^{-9} $~eV (from Eq.~(\ref{eq: axion mass})) which solves the strong CP problem. Furthermore, the PQ symmetry breaking takes place before inflation ($ f_a \gg T_{\rm reh} $) and never gets restored after inflation. In this circumstance, the initial misalignment angle of the axion is uniform throughout the Universe. Hence, the axion sits in the same sub-domain with a tiny quantum fluctuation, which avoids the formation of the domain wall below the QCD scale~\cite{Ibe:2019yew}. Finally, this axion may be observed by the future axion experiment Heterodyne Superconducting Radio Frequency (SRF) with its resonant SRF scan, see Ref.~\cite{Berlin:2020vrk}.

 \section{Other open questions in fundamental physics}
\label{sec: Other open questions in fundamental physics}

In this section, we will provide a brief discussion of the potential for the TNH framework to address the open questions related to quantum gravity and the cosmological constant problem. However, we will limit our discussion to the scale-invariant inflation scenario, as the presence of scale invariance may offer a promising approach to addressing these issues. \\

\textbf{Quantum gravity theory:} In the scenario with scale-invariant inflation, UV completions of gravity where the low energy Einstein
gravity emerges from seem softened due to the fact that there is no graviton propagator in this theory until the Planck scale is dynamically induced. Such a UV completion of gravity has to be scale-invariant, while the theory might only contain a metric, matter fields with non-minimal couplings, general covariance and no curvature terms standing alone. A possible example may be found in Ref.~\cite{Shaposhnikov:2008xb}, where they have studied this theory with the action in Eq.~(\ref{eq: SI action 1}) in the case where the General Relativity is replaced by Unimodular Gravity. Moreover, a quantization of such a theory is discussed in Ref.~\cite{Smolin:2009ti}. However, implementation of this setup into our model requires a more detailed investigation of such a theory, which is beyond the scope of the present paper. \\ 

\textbf{The cosmological constant problem:} Another important aspect of scale invariance is the absence of scales, resulting in a vanishing cosmological constant in the classical action. However, the cosmological constant is required to be vanishingly small~\cite{Nojiri:2009kx,Nojiri:2010wj} to explain the observed accelerated expansion of the Universe, as established by bayonic acoustic oscillations (BAO)~\cite{Weinberg:2013agg}, Ia supernova observation~\cite{Lee:1972yfa,Lee:1973fn} and cosmoic microwave background (CMB) anisotropies~\cite{Planck:2018vyg}. In this scale-invariant model, this smallness may be explained if the contributions from the quantum effects (vacuum polarization) are weak enough.

After reheating in the scale-invariant inflation scenario, the fields can enter a second slow-roll phase which leads to an effective cosmological constant given by~\cite{Ferreira:2016wem} \begin{equation}\begin{aligned} \Lambda_{\rm eff}=\frac{\lambda_\phi/4+\lambda_H \mu^4/4+\lambda_{H\phi}\mu^2/2}{\alpha_\phi +\alpha_H\mu^2}\phi_0^2\,,\label{eq: eff cosmological constant} \end{aligned} \end{equation} where $ \mu \equiv \langle \sigma_{H,f}\rangle  / \langle \phi_f \rangle \ll 1 $. With above ordering of the coupling, $ \lambda_H \gg \lambda_{H\phi} \gg \lambda_\phi $, we have that $ \Lambda_{\rm eff}\leq \lambda_H \sigma_{H,f}^4/(4 m_P^2)  $. If this cosmological constant is non-zero, a second slow-roll phase will result in eternal inflation. To require zero cosmological constant, the couplings have to be fine-tuned such that the potential has the form of a perfect square (further discussed in Ref.~\cite{Ferreira:2016wem}). We left a more in-depth study of the naturalness of these couplings to future work.

\section{Summary and conclusions}
\label{sec: Summary}

In this work, we explore the potential of Technically Natural Higgs (TNH) models to tackle key questions in fundamental physics. Specifically, we focus on six challenges: electroweak (EW) naturalness problem, neutrino masses and flavour mixing, inflation, matter-antimatter asymmetry, dark matter (DM) and the strong CP problem.

Our findings reveal that TNH models, featuring an interplay between elementary and composite Higgs states at a much higher scale than the EW scale ($ f\gg v_{\rm EW} $), provide a promising solution to the EW naturalness problem. Traditionally, reaching such a high compositeness scale would demand an implausibly fine-tuned vacuum alignment. These models, as proposed in Ref.~\cite{Rosenlyst:2021tdr}, introduce a mechanism that softly breaks a global $ \mathbb{Z}_2 $ symmetry, dynamically driving EW symmetry breaking (EWSB) and fermion mass generation at scales as large as the Planck mass, effectively resolving the EW naturalness problem.

In Section~\ref{sec: A Concrete Partially Composite Higgs Model}, we demonstrated this mechanism in a minimal partially composite two-Higgs model with the $ \SU(6)/\SP(6) $ coset, featuring both $ \mathbb{Z}_2 $--odd and --even Higgs doublets as pseudo-Nambu-Goldstone bosons (pNGBs). Notably, the $ \mathbb{Z}_2 $--odd scalar doublet triggers EWSB and charged fermion mass generation based on a small vacuum misalignment ($ \sin\theta \ll 1 $), leading to softly breaking the $ \mathbb{Z}_2 $ symmetry of the composite dynamics. Furthermore, as shown in Section~\ref{sec: Loop-Induced Neutrino Masses}, a natural near-degeneracy of the neutral components of the $ \mathbb{Z}_2 $--even scalar doublet features small loop-induced neutrino masses, where the composite dynamics generates this non-degeneracy via the EW gauge and Yukawa interactions. Conversely, in the fundamental scotogenic model, achieving such a near-degeneracy requires a very small scalar coupling~\cite{Ma:2006km}.

In Section~\ref{sec: RG analysis and vacuum stability}, we analyzed vacuum stability for this TNH model across different compositeness scales, ranging from TeV to Planck mass. Figure~\ref{fig: RG running without with neutrinos} shows phase diagrams illustrating stability for scenarios with and without scotogenic neutrinos. When $\Lambda_{\rm HC} \approx 4\pi f = m_{P}$ with scotogenic neutrinos, only a metastable vacuum is possible. For smaller $f$ within $10^{9}~\text{GeV} \lesssim f \lesssim 10^{14}$GeV, a stable vacuum is attainable. However, a smaller decay constant $ f $ introduces a new naturalness challenge, as it necessitates $m_H \lesssim \mathcal{O}(f)$ to prevent instability. With both a light 125 GeV Higgs boson and a heavy Higgs state around $\mathcal{O}(f)$, we can observe the realization of various inflation scenarios within this model. Remarkably, this involves the radiative induction of a bare Higgs mass parameter, $m_H\approx f$, within the range of $10^9~\text{GeV} \lesssim m_H \lesssim 10^{14}$ GeV.

Additional investigations in Section~\ref{sec: The inflationary paradigm} delve into these diverse inflation scenarios, encompassing the original Higgs inflation, Starobinsky $ R^2 $ inflation and scale-invariant inflation. These scenarios typically suffer from hierarchy problems arising from radiative corrections to the Higgs mass parameter. Within the TNH framework, these corrections do not jeopardize the lightness of the SM-like Higgs boson, as the generation of large mass parameters remains technically natural. In Sections~\ref{sec: Leptogenesis}--\ref{sec: QCD axion dark matter}, we have also examined possibilities for answering the remaining open questions. Thermal leptogenesis is explored as an explanation for the matter-antimatter asymmetry, where CP asymmetry from right-handed neutrino (RHN) decays contributes to the observed baryon-to-photon ratio. Additionally, we propose a scenario for explaining DM through the existence of a composite super-heavy DM (SHDM) particle. For scale-invariant inflation, the angular component of the inflaton may solve the strong CP problem and provide DM by becoming a QCD axion DM candidate.

In the following, we will provide a summary of the conclusions drawn from this study, focusing on the extent to which TNH models within the three inflation scenarios can offer solutions to the six fundamental physics problems: \\

\textbf{(i) Original Higgs inflation:} This inflationary scenario restricts the Higgs bare mass to the range of $1\times 10^{12}\text{ GeV} \ll m_H \lesssim 3\times 10^{16}\text{ GeV}$, necessitating the existence of a stable Higgs vacuum. Including neutrinos (lower panel in Figure~\ref{fig: RG running without with neutrinos}), viable Higgs inflation is feasible at $f=m_H \sim 10^{14}$~GeV and $t_\beta > 10$, with reheating at around $10^{14}$~GeV~\cite{Aoki:2022dzd}. This example requires $M_1 \ll f$ (as depicted in Figure~\ref{fig: f vs ml leptogenesis}) to produce the observed baryon asymmetry of the Universe (BAU), because of the high value of $f = m_H \gg 10^{12}$~GeV, which leads to small neutrino Yukawa couplings ($h^{ij} \ll 0.01$). Hence, generating a sufficiently small asymmetry through thermal leptogenesis without the need for fine-tuning presents a challenge. On the other hand, a composite SHDM with approximately $4\times 10^{14}$~GeV mass (comprising identical hyper-fermion masses) emerges as a promising solution for DM. 

In summary, within this inflation framework, the TNH model potentially addresses four out of the six issues for $f \sim 10^{14}$~GeV. However, challenges persist in solving the BAU and strong CP problems. Moreover, the conventional setup of the original Higgs inflation requires a large non-minimal coupling to the Ricci scalar, $ \alpha_H \sim 10^4 $, leading to a cutoff scale significantly below the Planck scale, $ \Lambda \lesssim m_P/\alpha_H \ll m_P $~\cite{Burgess:2009ea,Barbon:2009ya,Burgess:2010zq,Lerner:2009na,Park:2018kst}, which gives rise to the unitarity problem. This prompts exploration of alternative inflation scenarios. \\

\textbf{(ii) Starobinsky $ R^2 $ inflation:} In this inflationary paradigm, the loop-induced Higgs bare mass is approximately given by $ m_H \approx (5.5\times 10^7~\text{GeV}) \sqrt{\alpha_H (\alpha_H-3)} $. Within the range of $ 0.06 < \alpha_H < 11-12 $, the vacuum can exist in a metastable state, resulting in a Higgs bare mass of $ m_H < (5-6)\times 10^8~\text{GeV} $. This stands in contrast to the traditional Higgs inflation framework, where a stable vacuum is required. Additionally, the reheating temperature for this scenario is estimated to be approximately $ 10^{13} $~GeV, as reported in Ref.~\cite{Aoki:2022dzd}. When $ f $ falls within the range of approximately $ (5-6)\times 10^8 $~GeV or $ 3\times 10^9 \lesssim f \lesssim 1\times 10^{12} $~GeV, thermal leptogenesis successfully generates the observed BAU. Additionally, a composite SHDM candidate with a mass of approximately $ m_\Phi \sim 9\times 10^{9} $~GeV can account for the observed DM relic abundance through gravitational interactions. 

Altogether, this TNH model, incorporating Starobinsky $ R^2 $ inflation, has the potential to address five out of the six open questions. This can be achieved with $ f < (5-6)\times 10^8 $~GeV (resulting in a metastable vacuum and $ t_\beta \approx 5 $) or in the range $ 3\times 10^9 \lesssim f \lesssim 1\times 10^{12} $~GeV (yielding a stable vacuum and $ t_\beta \gtrsim 9 $). Finally, we explored the scale-invariant extension. \\

\textbf{(iii) Scale-independent $ R^2 $ inflation:} For scale-invariant inflation, radiative mass parameters emerge for the elementary weak doublets, $ H $ and $ H_{\nu} $, with approximate values of $ m_{H,H_\nu} \approx (5\times 10^7~\text{GeV})\alpha_H $. This aligns with the scale-dependent $ R^2 $ inflation scenario. However, the introduction of mixing terms such as $H^\dagger H \vert\phi\vert^2 $ and $H_\nu^\dagger H_\nu \vert\phi\vert^2 $ results in larger mass parameters, thereby expanding the parameter space, as illustrated in Figure~\ref{fig: mH alphaH}.

The lower panel of Figure~\ref{fig: RG running without with neutrinos} demonstrates that favorable parameter points, leading to either stable or metastable vacua, are predominantly clustered around $ f=\mathcal{O}(10^{12}) $~GeV. These scenarios lead to both neutrino mass generation and resolution of the EW hierarchy problem up to this scale. Notably, these neutrino masses are naturally suppressed due to a near-degeneracy of the neutral components of the scalar doublet $ H_\nu $, arising from the composite dynamics through the EW gauge and Yukawa interactions (refer to Eq.~(\ref{Eq: MR MI mass matrices})). In contrast, the fundamental scotogenic model proposed in Ref.~\cite{Ma:2006fn} relies on an extremely small scalar coupling to achieve a similar near-degeneracy. Indeed, this interesting scale of approximately $ f=\mathcal{O}(10^{12}) $~GeV can be loop-induced, particularly by introducing scale-invariant inflation. Coincidentally, this is achievable via a non-minimal coupling $ \alpha_H $ of order unity, as depicted in Figure~\ref{fig: mH alphaH}. Moreover, given that neutrino masses have already been radiatively generated through the scotogenic mechanism, it becomes plausible to account for the observed BAU via leptogenesis during the early Universe. Importantly, this approach does not necessitate additional ingredients but is intriguing in that it requires RHN masses to be close to the interesting scale, as shown in Figure~\ref{fig: f vs ml leptogenesis}.

Finally, an elegant and minimal solution to both the DM puzzle and the strong CP problem involves considering the angular component of the inflaton field, $ \phi $, as a QCD axion, thereby identifying it as a candidate for DM. This interpretation can be realized with the special compositeness scale $ f=\mathcal{O}(10^{12}) $~GeV and an initial misalignment angle of $ 0.01-0.1 $. Consequently, the axion decay constant is determined as $ f_a = v_\phi \sim 10^{15} $~GeV, where $ v_\phi $ is the VEV of $ \phi $. Remarkably, $ \phi $ serves a dual role as both inflaton and QCD axion DM, while also contributing to the emergence of the Planck mass through scale symmetry breaking. Furthermore, through Yukawa and scalar couplings, the VEV $ v_\phi $ generates masses for heavy particles ($ m_i, M_i, m_{H, H_\nu} \sim \mathcal{O}(10^{12}) $~GeV and $ m_Q \sim \mathcal{O}(10^{15}) $~GeV), yielding an economically appealing framework. Thus, $ \phi $ plays an additional role by also providing the mass-dependent parameters (approximately $ \mathcal{O}(10^{12}) $~GeV) within the high-scale partially composite dynamics, while the underlying composite dynamics dynamically explains the hierarchy between this compositeness scale and the EW scale, as well as accounting for the small neutrino masses and the observed BAU. \\

In conclusion, this paper illuminates the significance of an intriguing scale, approximately at $ \mathcal{O}(10^{12}) $~GeV, within the minimal TNH framework. For the scale-invariant version of the minimal TNH model, this scale plays a vital role in effectively addressing all of the aforementioned challenges in particle physics and cosmology.

\section*{Acknowledgements}
I acknowledge partial funding from The Independent Research Fund Denmark, grant numbers DFF 1056-00027B. I would like to thank P.~G.~Ferreira, C.~T.~Hill and P.~Sørensen for discussions. 

\bibliography{TNH.bib}
\bibliographystyle{JHEP}

\end{document}